\input amstex
\magnification=1200
\documentstyle{amsppt}
\NoBlackBoxes
\nologo
\pagewidth{6.5truein}
\pageheight{9.0truein}
\def\On{\Omega_{\nu}}
\def\mn{m_{\nu}}
\def\pn{\psi_{\nu}}
\def\Sn{S_{\nu}}
\def\Bn{B_{\nu}}
\def\Cn{C_{\nu}}

\def\Un{U_{\nu}}
\def\Mn{M_{\nu}}
\def\pz{z\frac{\partial}{\partial z}}
\def\px{\frac{\partial}{\partial x}}
\def\dx{\frac{d}{dx}}
\def\ad{\operatorname{ad}}
\def\inv{^{-1}}
\def\pd#1#2{\frac{\partial{#1}}{\partial{#2}}}
\def\tr{\operatorname{tr}} 
\def\ss{\vskip .3cm}
\def\ms{\vskip .5cm}
\def\a{\alpha}
\def\b{\beta}
\def\lmat{\left[\matrix}
\def\rmat{\endmatrix\right]}
\def\tl{T_{\lambda}}
\def\l{\lambda}
\def\L{(L^{k/n})_+}
\def\Lz{\Lambda_z}

\topmatter
\title Integrable Systems and Isomonodromy Deformations\endtitle
\rightheadtext{}
\author Richard Beals
and David H. Sattinger\endauthor

\affil Yale University and University of Minnesota\endaffil
\leftheadtext{}
\footnote[]{Research of the authors was
supported by National Science
Foundation grants DMS-8916968 and DMS-9123844}

\abstract{We analyze in detail three classes of isomondromy
deformation problems associated with integrable systems. The
first two are related to the scaling invariance of the
$n\times n$ AKNS hierarchies and the Gel'fand-Dikii hierarchies.
The third arises in string theory as the representation of the
Heisenberg group by $[(L^{k/n})_+,L]=I$ where $L$ is an $n^{th}$
order scalar differential operator. The monodromy data is
constructed in each case; the inverse monodromy problem is
solved as a Riemann-Hilbert problem; and a simple proof
of the Painlev\'e property is given for the general case.}
\endabstract

\endtopmatter
\document
\baselineskip 15pt
Introduction

1. Overdetermined systems and isomonodromy equations

2. The forward monodromy problem at $z=\infty$

3. The forward monodromy problem at $z=0$

4. The inverse problem and the Painlev\'e property

5. Rational solutions

6. B\"acklund transformations

7. Scaling, self-similarity, and construction of isomonodromy
deformations

8. Gel'fand-Dikii equations and isomonodromy 

9. Isomonodromy deformations and string equations 
\ms
{\bf Introduction}
\ss
It has been known since work of Ablowitz and Segur [AS1] that there is an intimate relationship
between equations like KdV which are integrable by the inverse scattering
method  and  Painlev\'e equations; see also [AS2].
It was observed by Flaschka and Newell [FN] that just as KdV gives an 
isospectral flow for the Schr\"odinger
operator, Painlev\'e equations are monodromy preserving flows for
linear systems with irregular singular points. 
Certain of these problems have been investigated in detail, for $2\times 2$
systems and second order scalar problems: [FN],
[FZ], [IN], [JMU], [JM1], [JM2].

In this paper we discuss general isomonodromy problems
associated to $n\times n$ systems and higher order equations.  The
corresponding isomonodromy equations are generally of order greater
than 2, so they are not the classical Painlev\'e transcendents.
However they do have the Painlev\'e property,
in fact the stronger property that any solution has a single-valued
meromorphic extension to the entire plane; cf. [Ma], [Mi].
We hope, among other things,
to simplify and clarify the treatment of isomonodromy deformations
and their relation to Riemann-Hilbert problems, on the principle that 
the more general the case the less reliance on special features.

The paper is organized as follows.  In \S 1 we describe the formal
connection between matrix isomonodromy equations and overdetermined
$n\times n$ systems in two variables.  This connection is made rigorous
in the next three sections, which describe the forward problem at the
singular points $z=\infty$ and $z=0$, connect it to a Riemann-Hilbert
problem, and prove the Painlev\'e property.  A few examples of equations
and systems which occur in this context are: the Painlev\'e II equation
$$
4(xu)_x+u_{xxx}-6u^2u_x=0;
$$
the system of three equations of order one
$$
(xu_i)_x=a_i u+ b_i u_ju_k,\qquad \{i,j,k\}=\{1,2,3\};
$$ 
the system of two equations of order two with cubic nonlinearity
$$ 
(xu_1)_x+\tfrac12 u_{1xx}-u_1^2 u_2=0=(xu_2)_x-\tfrac12 u_{2xx}+u_2^2 u_1;
$$
and the system of two equations of order two with quadratic nonlinearity
$$
(xu_1)_x+\tfrac{i}{\sqrt3} u_{1xx}+2u_2u_{2x}=0=(xu_2)_x-
\tfrac{i}{\sqrt3} u_{2xx}+2u_1u_{1x}.
$$

In \S 5 the  rational solutions of the isomonodromy equations
are constructed by solving finite linear systems.  This extends 
results of Airault [Ai], who found B\"acklund transformations giving
rational solutions of some Painlev\'e equations.  We develop   
the gauge theory of B\"acklund transforms in \S6. The
gauge transformations take the wave functions for one solution
to those of a new solution and thus transform solutions to solutions.

Isomonodromy deformations arise from integrable systems in two ways.
Some can be obtained as {\it self-similar solutions} of the given nonlinear
evolution equations; see [AS2]. 
This construction is given in \S 7 for isospectral deformations
of $n\times n$ first-order operators (AKNS-ZS systems); each of the 
examples above is of this type.  In \S 7 we also treat 
the isospectral deformations of an $n$-th order scalar
differential operator (the Gel'fand-Dikii hierarchy).  Examples include
the equation
$$
(xu)_x+u+\tfrac14u_{xxx}+\tfrac32 uu_x=0
$$
which corresponds to self-similar solutions of the KdV equation and
the system
$$
(xu_1)_x=u_{1xx}-2u_{0x},\quad (xu_0)_x+u_0=\tfrac23u_{1xx}+\tfrac23
u_1u_{1x}-u_{0xx}
$$
which corresponds to self-similar solutions of the Boussinesq system.
In all these cases
the Lax pair for the integrable system can be rescaled to obtain a
Lax pair for the corresponding isomonodromy deformation problem.
This was done in detail in the $2\times 2$ case by Flaschka and 
Newell [FN] for the
modified KdV equation and its associated isomonodromy problem,
the Painlev\'e II equation, as well as the sine-Gordon equation and
the Painlev\'e III equation  (cf. also Its and Novokshenov [IN]).
In section 8 we show that all self-similar solutions of Gel'fand-Dikii
flows are in fact solutions of isomonodromy equations, and we treat 
the direct and inverse
problems for these equations.  The results are analogous to those
of \S\S 2, 3, 4. 

A second class of isomonodromy problems was obtained by M. Douglas
[Do] in recent two-dimensional theories of quantum gravity.
These problems are obtained by replacing $\dot L$ by $\hbar I$ in the
Lax equation $\dot L=[\,(L^{k/n})_+\,,\,L\,]$ for the Gel'fand-Dikii
flows.  This device leads to a representation of the Heisenberg
group and to a class of isomonodromy problems different from those
obtained by the scaling invariance.  Two of the simplest examples are
$$\align
&\tfrac14 u_{xxx}+\tfrac32 uu_x+\hbar=0;\\
&\tfrac1{16}u^{(5)} + \tfrac58 uu_{xxx}+\tfrac54u_x u_{xx}+\tfrac{15}8 
u^2u_x+\hbar=0.\endalign
$$
This class of isomonodromy problems was treated by Moore [Mo1], [Mo2].
We thank him for useful discussions of the topic. In \S 9 we focus on
some aspects of the mathematical analysis of these problems and give a
different, more complete and self-contained treatment.
\ms
\bf 1. Overdetermined systems and isomonodromy equations \rm
\ss
In this section we outline the formal connection between certain
overdetermined linear $n\times n$ systems and monodromy preserving
equations. 

Let $J$ and $\mu$ be diagonal matrices belonging to the space $M_d(\Bbb C)$
of $d\times d$ complex matrices; {\it we assume that each has distinct diagonal
entries and has trace zero.}  Suppose that $q=q(x)$ is an off-diagonal 
matrix-valued function
defined on some real interval or some connected complex domain $I$.
Consider the overdetermined system for a matrix-valued function
$\psi(x,z)$ of real or complex $x$ and complex $z$, $\psi\in SL(n,\Bbb C)$,
$n\geq 2$:

$$\align
\frac{\partial\psi}{\partial x}& = [zJ + q(x)]\psi\,; \tag1.1 \\ 
z\frac{\partial\psi}{\partial z}& = A(x,z)\psi =\sum^n_{j=0}z^j A_j(x)\psi\,,
\qquad A_n\equiv \mu\,.
\tag1.2 \endalign
$$
The compatibility condition (``zero curvature condition'') for these equations
is
$$
\bigl[\,\px - (zJ+q)\,,\,\pz - A(x,z)\,\bigr]=0.\tag1.3
$$
Set
$$
A_j = F_{n-j},\ \ n>1;\qquad A_1 = F_{n-1}+xJ;\qquad
A_0 = F_n + xq(x).\tag1.4
$$
Then the zero-curvature condition (1.3)
is equivalent to the sequence of conditions
$$
 F_0 = \mu,\qquad [\,J\,,\,F_{j+1}] = \bigl[\,\frac{d}{dx} - q\,,\,F_j\,\bigr],\quad 0\leq j<n,
\tag1.5  
$$
together with the equation 
$$
\bigl[\,\frac{d}{dx}- q\,,\,xq + F_n\,\bigr]= 0\tag1.6
$$
	
The conditions (1.4) do not determine the $F_j$ uniquely but, as we shall
indicate,
there is a determination for which the $F_j$ are polynomials in $q$ and its
derivatives.  Then (1.6) is a nonlinear differential equation for $q$, and
we show that it is an \it isomonodromy equation \rm for the system
of differential equations in the complex domain
(1.2): under the flow (1.6), the monodromy of (1.2) is invariant. 

\noindent{\bf Examples:}  1. The simplest nontrivial example is obtained 
with $n=2$, $k=3$, $\mu=J=\text{diag}(1,-1)$, and
$$
q(x)=u(x)\lmat 0&1\\1&0\rmat\,.
$$
Then
$$
F_1=q,\quad F_2=\frac12\lmat -u^2 & u_x\\ -u_x & u^2\rmat,\quad
F_3=\frac14\lmat 0 & u_{xx}-2u^3\\ u_{xx}-2u^3 & 0\rmat,
$$
and equation (1.6) becomes the equation satisfied by self-similar
solutions of the mKdV equation, namely the Painlev\'e II equation
$$
4(xu)_x+u_{xxx}-6u^2u_x=0.
$$
See \S\S 7, 8 for a full discussion of self-similarity.

2. Although we assume generally that $n$ is at least two, certain cases
with $n=1$ can be treated, e.g when $J$ and $\mu$ are both real.
If we assume this and assume that $q+q^t=0$, then the $3\times 3$ case 
with $n=1$ and $k=1$ leads to a system of equations of the form
$$
(xu_i)_x + a_i u + b_i u_j u_k = 0,\qquad \{i,j,k\}=\{1,2,3\},
$$
where the $a_i$'s and $b_i$'s are constants.  These equations correspond
to self-similar solutions of the three-wave interaction equation.
Some more examples are sketched at the end of this section.
\ss

We suppose from now on that $q$ is a smooth function on the domain $I\subset
\Bbb R$ and that it is off-diagonal:
$$
q(x)_{jj} = 0,\qquad 1\leq j\leq d.
$$
Let
$$
\psi(x,z)=m(x,z)e^{\Phi(x,z)},\qquad \Phi(x,z)=\frac1{n}z^n\mu+xzJ.\tag1.7
$$
The equations (1.1), (1.2) for $\psi$ are equivalent to 
$$\align
\pd{ m}{x} &= [\,zJ\,,\,m\,]+qm\,, \tag1.8\\
z\pd{ m}{z} &= A(x,z)\,m - m\,(z^n\mu+xzJ)\,.\tag1.9\endalign
$$
By a {\it formal solution\/} of (1.8) we mean a formal power series in
$z\inv$, 
$$
m(x,z)=\sum^\infty_{j=0}z^{-j}f_j(x),\ \ \ \ f_j\in 
C^\infty(I; M_d(\Bbb C)),\quad f_0 = \pmb 1\,
$$
which satisfies (1.8) formally; this means that for each $N$
$$
\left(\px-z\ad J- q\right)\sum_{j=0}^N z^{-j} f_j(x) = O(z^{-N}).
$$
Equivalently,
$$
\frac{df_j}{dx}-qf_j = [\,J\,,f_{j+1}\,]\,.\tag1.10
$$
These relations determine the off-diagonal part of $f_{j+1}$ from $f_j$
by inverting $\ad J$.  
At the next step they determine the diagonal part of $f_{j+1}$, up to
a constant matrix, in terms of the off-diagonal part of $f_{j+1}$ since
the diagonal part of $[\,J\,,\,f_{j+2}\,]$ vanishes and the diagonal part
of $qf_{j+1}$ involves only the off-diagonal part of $f_{j+1}$.  Thus
these relations are solvable recursively, so formal solutions
exist.  Note in particular that $q=f_1J-Jf_1$ and that the diagonal part
of $f_1$ may be taken to be $0$.  For later use we need to go one step
further in this discussion.  Let us fix a point $x_0$ and choose the
unique formal solution with the property that $f_j(x_0)$ is off-diagonal
for all $j>0$.  Then one can prove by induction that each
Taylor coefficient at $x_0$ of each $f_j$ is given by a universal polynomial
in the Taylor coefficients at $x_0$ of the entries of $q$.  Moreover,
each of these polynomials has constant term $0$.  {\it In particular, 
$f_{j+1}(x_0)$ is a polynomial in the $j$-jet of $q$ at $x_0$.}

We define formal solutions of (1.9) in an analogous manner.
 
\proclaim{\bf Theorem 1.1} Let $m=\sum^{\infty}_{j=0}z^{-j}f_j(x)$ 
be a formal solution of (1.8). The formal series
$$
F = m\mu m^{-1} =\sum^{\infty}_{j=0}z^{-j}F_j(x) \tag1.11
$$
is independent of the choice of $m$.  The coefficients $F_j=F_{j,\mu}$ are 
traceless polynomials in $q$ and its derivatives:
$$ 
F_j(x) = P_j(q(x), q'(x),\dots,q^{j-1}(x)) .\tag1.12
$$
Moreover, the $F_j$ satisfy the conditions (1.5).

If $m$ is also a formal asymptotic solution of (1.2), then the relation
between the coefficients $A_j$ and $F_j$ is given by (1.4) and $q$ 
satisfies the isomonodromy equation (1.6).
\endproclaim
\demo{Proof} 
Suppose that $m_1$ and $m_2$ are two formal asymptotic solutions; then $m_2$
has a formal inverse and $g=m_1^{-1}m_2$ is a formal asymptotic solution of
the equation
$dg/dx = [zJ,g]$.  The corresponding relations are 
$$
[J,~g_{j+1}] = \frac{dg_j}{dx},
$$
and since $g_0=\pmb 1$ it follows recursively that each $g_j$ is diagonal
and constant. Therefore $m_2$ is a formal power series product
$m_2 = m_1g$ with $g$ diagonal and so $m_2\mu m_2^{-1}= m_1\mu m_1^{-1}$.

It is clear from the preceding that $m\mu m\inv$ is independent of the 
choice of formal solution $m$.  The assertion about the coefficients
of $m\mu m\inv$ may be derived by localizing and using a result of [Sa],
but we give here another proof which adapts more readily for use in \S 8.
Fix any $x_0$ in the domain of $q$, and let $m$ be the formal solution
whose diagonal part at $x_0$ is $\pmb 1$.  From the remarks above we
conclude that the coefficients of $m\mu m\inv$ at $x_0$ are given by
universal polynomials in the entries of $q$ and their derivatives at
$x_0$.  But this fact is independent of the choice of $m$ and of $x_0$.   

If $m$ is a formal solution of (1.8), then $F=m\mu m\inv$ is 
readily seen to be a formal solution of 
$$
\bigl[\,\frac{d}{dx} - zJ - q\,,\,F\,\bigr] = 0\,.\tag1.13
$$
Therefore its coefficients satisfy the recursion relations corresponding
to (1.13), which are (1.5).

Finally, suppose that $m$ is also a formal solution of (1.9).  Then
 $zm_z$ has no constant term, so we conclude that 
(as formal power series in $z\inv$)
$$\align
A(x,z)&=m(x,z)\,[z^n+xzJ]\,m(x,z)\inv + O(z\inv)\\
&= z^nF + xzJ + xf_1J-xJf_1+O(z\inv)= z^n F + xzJ + xq +O(z\inv)\,.\endalign
$$
This is equivalent to (1.4). Now (1.3) can also be obtained by equating
terms in formal power series expansions, and we deduce the isomonodromy
equation (1.6).  
\qed\enddemo
Let us note explicitly that 
$$\align
F_0 &= \mu\,,\quad F_1 =(\ad J)\inv[\,\mu\,,\, q\,]\,,\\
 F_{j+1}&=\left(\frac{d}{dx}\right)^j(\ad J)^{-j}[\,\mu\,,\,q\,]+
\{\,\text{terms of order}\ <j\,\}.\endalign
$$ 
We assume from now on that the $F_j$ are those in Theorem 1.2.

\proclaim{\bf Corollary 1.2} The isomonodromy equation (1.6) is an algebraic
differential equation for the matrix function $q$.\endproclaim

\noindent{\bf Further examples:} 3. Here we take $n=2$, $J=\mu=
\text{diag}(1,-1)$, and $k=2$, with $q$ a general off-diagonal matrix,
$$
q=\lmat 0 & u_1\\ u_2 & 0\rmat.
$$
Some computation shows that $F_1=q$, $F_2=\tfrac12 qJq+\tfrac12 Jq_x$, 
and equation (1.6)
becomes $(xq)_x +\tfrac12 Jq_{xx}-q^2Jq=0$.  Explicitly,
$$
(xu_1)_x+\tfrac12 u_{1xx}-u_1^2 u_2=0=(xu_2)_x-\tfrac12 u_{2xx}+u_2^2 u_1.
$$
If we replace $J$ by $iJ$ the formula is the same and is compatible with
the reductions $u_2=\pm\bar u_1$, in which case the solutions are exactly
the self-similar solutions of the cubic nonlinear Schr\"odinger equations.

4. Finally we take $n=3$, $J=\text{diag}(\a,\a^2,1)$, where $\a=e^{2\pi i/3}$
 is a primitive cube root of $1$, $\mu=J^2$, and $k=2$.  
We assume that $q$ has
the form $q=u_1\Pi + u_2\Pi^2$, where $u_1$ and $u_2$ are scalar and $\Pi$
is the permutation matrix
$$
\Pi=\lmat 0 & 1 & 0\\0 & 0 & 1\\1 & 0 & 0\rmat.
$$
Then $F_1$=$[J^2,r]$ and $F_2=Jr_x+r_xJ+q^2-2u_1u_2\pmb 1$, where
$$
r=(1-\a)\inv u_1 J^2\Pi+(1-\a^2)^{-1}u_2 J\Pi^2.
$$
Equation (1.6) becomes $(xq)_x+Jr_{xx}+r_{xx}+(q^2)_x-2(u_1u_2)_x\pmb 1$,
or explicitly the coupled quadratic nonlinear system
$$
(xu_1)_x+\tfrac{i}{\sqrt3}u_{1xx}+2u_2u_{2x}=0=(xu_2)_x-\tfrac{i}{\sqrt3}
 u_{2xx}+2u_1u_{1x}.
$$

\vskip .5cm
\bf 2. The forward monodromy problem at $z=\infty$\rm
\vskip .3cm
In this section we continue to assume that $q$ is a smooth, off-diagonal 
matrix function on a real interval or a connected complex domain $I$
and we assume it satisfies the algebraic differential equation (1.6). 
(Consequently $q$ is analytic in $I$.)  

Using a partial sum of a {\it formal} solution of the
$x$-equation as a gauge transformation (or parametrix) we show that the 
$z$-equation has {\it actual} 
solutions which are well-behaved in specified sectors and have
a special form.  Moreover, the classical Stokes matrices, which link the
solutions from different sectors, are independent of $x$.  (It is
classical to use
a formal solution of the $z$-equation as a gauge transformation to obtain
a tractable integral equation for the transformed solution;
cf. Chapter 5 of [CL].  The special feature here is that the phase function
which occurs in the $z$-asymptotics of the solutions has a very special
form, and this is a consequence of the fact that the formal solutions are
also formal solutions of the $x$-equation.)

Let the polynomial part of the formal series $z^nF$ be denoted 
$$
[z^n F]_+ = \sum^n_{j=0} z^{n-j}F_j\,.
$$
We assume that $q$ satisfies the isomonodromy equation (1.6):
$$
\bigl[\,\frac{d}{dx}-q\,,\,xq+F_n\,\bigr] = 0\,.
$$
In view of the recursion relations for the $F_j$, an alternative form is
$$
(xq)_x + [\,J\,,\,F_{n+1}\,]=0.
$$
It is convenient to introduce a condensed notation for the differential
operators which recur throughout:
$$
D_x\equiv \frac{\partial}{\partial x}-(zJ+q);
\qquad D_z\equiv z\frac{\partial}{\partial z}-[z^nF]_+ -x(zJ+q)
$$
Then the conditions (1.5) are satisfied and (1.6) is the remaining
requirement for the commutator condition
$$
[\,D_x\,,\,D_z\,]=0.\tag2.1
$$

We shall show that (1.2) has certain normalized solutions with asymptotic
expansions in sectors of the $z-$plane.  To describe these sectors, we set
$$
\Sigma(j,k) = \{z\in \Bbb C: \operatorname{Re}z^n(\mu_j-\mu_k) = 0\},\quad
1\leq j<k\leq d;\qquad\Sigma = \bigcup_{j<k}\Sigma(j,k).
$$
Each $\Sigma(j,k)$ is a union of $n$ lines through the origin, with angle
$\pi/n$ between successive rays.
The distinguished open sectors $\{\Omega_{\nu}\}$ which
we consider are described as follows.  Each sector is bounded by two rays
from $\Sigma$, and its interior contains exactly one ray from each 
$\Sigma(j,k)$.  We take the \it leading edge \rm of a sector to be the
boundary ray with greater argument 
and we number sectors in order of increasing argument of leading edge, from
some starting point.

\proclaim{\bf Theorem 2.1} Suppose that $q$ satisfies the isomonodromy 
equation (1.6).
Then for each $x$ and each sector $\Omega_{\nu}$ there is a unique solution
of
$$
z\frac{\partial\psi}{\partial z} = ([z^nF]_+ + xzJ + xq)\psi ,\quad z\in
\Omega_{\nu}\tag2.2
$$
which has the form $\psi_\nu=m_\nu e^{\Phi}$ as in (1.7), 
where the $m_{\nu}$ have the (same) asymptotic expansion
$$
m_{\nu}\sim\sum^{\infty}_{j=0}z^{-j}f_j(x)\quad\text{as}~~z\to\infty,
\tag2.3
$$
with $f_0 = \pmb 1$.  

The expansion (2.3) is valid uniformly as $z$ tends to
infinity in any closed subsector of $\Omega_{\nu}\cup\{0\}$ and as 
$x$ varies in any compact set.

There are constant matrices (Stokes matrices) $S_{\nu}$ such that the relation
between successive solutions is
$$
\psi_{\nu+1}(x,z) = \psi_{\nu}(x,z)\Sn,\qquad z\in\Omega_{\nu}\cap\Omega
_{\nu+1}.\tag2.4
$$

Finally, the $\psi_{\nu}$ satisfy 
$$
\det\pn \equiv 1\,;\quad\ 
\frac{\partial\pn}{\partial x} = (zJ+q)\pn.\tag2.5
$$\endproclaim 
 
The rest of this section is devoted to the proof of Theorem 2.1.  We begin with
a gauge transformation to convert (2.2) to a more tractable form for large
$z$.  Let $m = \sum z^{-j}f_j$ be a formal solution of (1.8) 
with $f_0=\pmb 1$ and set
$$
f(x,z) = \sum^n_{j=0}z^{-j}f_j(x).
$$
Then $f\mu f^{-1} = F + O(z^{-n-1})$, while (1.8) implies that 
$(zJ+q)f = zfJ+O(z\inv)$, so
$$
([z^nF]_+ +xzJ +xq)\,f = f\,(z^n\mu + xzJ) + O(z^{-1})\,.\tag2.6
$$

If we look for a solution of (2.2) in the form $\psi = f\hat\psi$, then
(2.2) is equivalent to 
$$
\frac{\partial\hat\psi}{\partial z} = (z^{n-1}\mu + xJ)\,\hat\psi + r(x,z)
\,\hat\psi
$$
with $r(x,z) = O(z^{-2})$.  Setting $\hat\psi = \hat me^{\Phi}$,
we convert this to the integral equation
$$
\hat m = \pmb 1 + \int^z_\infty e^{\Phi(z)-\Phi(\zeta)}r(\zeta)\hat m(\zeta)
e^{-\Phi(z)+\Phi(\zeta)}\,d\zeta,\tag2.7
$$
where we have suppressed the dependence on $x$.  
Note that (2.7) is a matrix equation and the path of integration may
differ entry by entry.

\proclaim{\bf  Lemma 2.2} In each sector $\Omega_{\nu}$, the paths of integration
in (2.7) may be chosen so that there is a solution for large $z$ in each 
closed subsector, with an asymptotic expansion valid uniformly in the
smaller sector.\endproclaim
\demo{Proof} Let
$$
\Sigma_{xz}(j,k) = \{\,\zeta\in\Bbb C:\operatorname{Re}[\Phi(x,z)-\Phi(x,\zeta)
]_{jj} =\operatorname{Re}[\Phi(x,z)-\Phi(x,\zeta)]_{kk}\,\}\,, \quad j<k.
$$
This set consists of $n$ regular curves, one of which goes through $z$. For
$z$ not in $\Sigma = \bigcup\Sigma(j,k)$ the curve through $z$ is asymptotic
to the nearest two rays of $\Sigma(j,k)$.  It follows that for each $z\in
\Omega_{\nu}$ we may choose the path of integration for the $(j,k)$ and
$(k,j)$ entries in (2.7) to lie along this branch and to be asymptotic to
the ray of $\Sigma(j,k)$ which lies in the interior of $\Omega_{\nu}$.
The effect is that conjugation by $e^{\Phi(z)-\Phi(\zeta)}$ multiplies
each entry by a \it bounded \rm exponential.  

If $z$ lies in a closed subsector of $\Omega_{\nu}$, the associated
paths of integration lie in $\{|\zeta|\geq R\}$ when 
$|z|$ is large.  Since
$r = O(z^{-2})$, it follows that eventually the integral equation (2.7) with
our choice of contour has a unique solution obtained by successive 
approximations. By a standard argument, this
solution has an asymptotic expansion which is valid uniformly in each
closed subsector. \qed\enddemo

Note that the solutions so obtained are holomorphic, and thus extend to the
entire sector; indeed they extend as  functions on
$\Bbb C\setminus 0$ (multi-valued in general).
 
\proclaim{\bf  Lemma 2.3} Suppose $\tilde\psi_{\nu}$ is a second solution of (2.2) in
$\Omega_{\nu}$, for fixed $x$, with the property that 
the limit of $\tilde\psi_{\nu}e^{-\Phi}$ as $z$ tends to infinity along
any ray in $\Sn$ is $\pmb 1$.
Then $\tilde\psi_{\nu} = \psi_{\nu}$.\endproclaim

\demo{Proof} There is a constant matrix $S$ such that 
$\tilde\psi_{\nu} = \psi_{\nu}
S$. From the asymptotic expansions it follows that the conjugation 
$$
e^{\Phi(x,z)}Se^{-\Phi(x,z)} = [\tilde\psi_{\nu}e^{-\Phi}]^{-1}
[\psi_{\nu}e^{-\Phi}]
$$
converges to $\pmb 1$ as $z$ tends to $\infty$ along any ray in 
$\Omega_{\nu}$.
Then the diagonal entries of $S$ are $1$. By boundedness,
$$
S_{jk}=0\quad\text{if}\quad \operatorname{Re}z^n(\mu_j-\mu_k)>0
$$
along such a ray.  By assumption, $\Omega_\nu$ contains such rays for each
$j\neq k$, so $S=\pmb 1$. \qed\enddemo

\proclaim{\bf  Lemma 2.4} The functions $\psi_{\nu}$ satisfy (2.5).\endproclaim

\demo{Proof} The commutativity property (2.1) implies that the function
$$
\tilde\psi_{\nu} = \bigl[\frac{\partial}{\partial x} - zJ - q
\bigr]\psi_{\nu}
$$
is also a solution of (2.2), so $D_x\tilde\psi_{\nu} = \psi_{\nu} T(x)$
for some matrix-valued function $T$.  
The construction of $\psi_{\nu}$ shows that the corresponding 
$m_{\nu}$ is differentiable with respect to $x$ and the  asymptotic
expansion of $m_{\nu}$ can be differentiated term by term with respect to
$x$.  Then
$$
m^{-1}_{\nu}\biggl(\frac{\partial}{\partial x} 
- z\operatorname{ad}J - q\biggl)m_{\nu}
= e^{\Phi}Te^{-\Phi}.\tag2.8
$$
The left side is bounded as $z\to\infty$ so,
as in the proof of Lemma 2.3,
 it follows that the matrix $T$ must be diagonal.  Thus the
right side of (2.8) is independent of $z$.  On the other
hand, a check of the asymptotics of the left side of (2.8) shows that the
leading term for large $z$ is $[m_{\nu 1},J]-q$, where $m_{\nu 1}$ is 
the coefficient of $z^{-1}$ in the asymptotic expansion of $m_{\nu}$.
This term is off-diagonal since $q$ is off-diagonal.  (This is the first
time we have used the assumption that $q$ is off-diagonal.)
Thus both sides of (2.8) must vanish, hence $D_x\psi_{\nu}
=0$.  We have proved in the process that
$$
q(x) = - [\,J\,,\,m_{\nu 1}\,],\quad\text{where}\quad m_{\nu}= 
1+z^{-1}m_{\nu 1}+ O(z^{-2}).\tag2.9 
$$
Finally, the equation (1.8) for $\mn$ implies that $\det\mn$ is constant,
hence is identically $ 1$.  We have assumed that $\Phi$ is traceless, 
so $\pn =\mn e^\Phi$ also has determinant $1$.
\qed\enddemo

\proclaim{\bf  Lemma 2.5} There are constant matrices $S_{\nu}$ such that for all 
$x$ in $I$
$$
\psi_{\nu+1}(x,z) = \psi_{\nu}(x,z)S_{\nu},\qquad  z\in\Omega_{\nu+1}\cap 
\Omega_{\nu}.\tag2.10
$$
\endproclaim
\demo{Proof} This is clear from the fact that in the common domain of
definition, both functions satisfy linear first order systems of differential
equations in $x$ and in $z$.\enddemo

The preceding lemma makes possible the final step in the proof of Theorem 2.1.

\proclaim{\bf  Corollary 2.6} The coefficients of the asymptotic series 
for $\mn$ do not depend on the sector.\endproclaim

\demo{Proof}  By Lemma 2.3,
$$
m_{\nu} = m_{\nu+1}e^{\Phi}S_{\nu}^{-1}e^{-\Phi}
$$
on any ray in $\Omega_{\nu+1}\cap\On$, and our arguments show that
the only non-zero entries of $S_{\nu}$ must remain bounded under the 
conjugation by $e^{\Phi}$ as $z$ tends to infinity on the ray.
Therefore the off-diagonal entries decay
exponentially as $z\to\infty$.  This implies that the asymptotic
expansions of $\mn$ and $m_{\nu+1}$ are the same.
\vskip .5cm

\bf 3. The forward monodromy problem at $z=0$ \rm
\vskip .3cm

To complete the analysis of the $z$-equation one must examine the
behavior at the regular singular point $z=0$.  

We continue to assume that $q$ is a solution of the isomonodromy
equation (1.6) on a domain $I$.  We take the sectors $\On$ and the
matrix functions $\pn$, $\mn$, $\Phi$ as in \S 2, and we assume that
there are $N$ sectors numbered cyclically: $\Omega_N = \Omega_0$.

\proclaim{\bf  Theorem 3.1} There are invertible constant matrices $C_{\nu}$
and a matrix function $U(z)$, holomorphic
and invertible in $\Bbb C\setminus 0$, such that
$$
\mn(x,z)e^{\Phi(x,z)}
C_{\nu}\inv U(z)\inv \equiv \mn(x,z)e^{\Phi(x,z)}U_{\nu}\inv
$$
is regular at $z=0$ for every $x\in I$ and is independent of $\nu$.
\endproclaim

\proclaim{\bf  Lemma 3.2} There is a fundamental solution of $D_x \psi = 0$,
$D_z\psi = 0$ which has the form 
$$
\psi(x,z)=f(x,z)V(z)z^B,\quad z\in\Bbb C\setminus 0,\tag3.1 
$$
with  $f(x,\cdot)$ entire, $V$ entire, $B$ constant, and $\det(Vz^B)\equiv 1$.
\endproclaim
\demo{Proof} Choose $x_0\in I$ and let  
 $f(x,z)$ be the unique solution of $D_xf=0$ which
satisfies the initial condition $f(x_0,z)=\pmb 1$; 
it is entire as a function of $z$. Since $[D_x,D_z]=0$,
it follows that $D_xD_zf=D_zD_xf=0$, so $D_zf$ is also in the kernel of $D_x$.
Therefore $D_zf=f(x,z)C(z)$ for some (entire) function $C$. 
Note that $\det f$ is constant with respect to $x$, hence is identically $ 1$. 
We look for $\psi$ in the form $\psi(x,z)=f(x,z)U(z)$.
The necessary  and sufficient condition for $D_z(fU) = 0$ is
$$z\frac{dU}{dz}= -CU. \tag3.2
$$
From the equation $D_zf = fC$ and the condition $f(x_0,z)=\pmb 1$ we find that
$C$ is a polynomial in $z$ and 
$-C(0) = x_0q(x_0) + F_n(x_0)$.  Equation (3.2) has a 
regular singular point at the origin, so it has a fundamental solution
$U(z) = V(z)z^B$, where V is entire and $B$ is constant [CL, Ch. 4,
Theorem 4.2]. Now $\det U$ is constant, so we may choose $\det U\equiv 1$.
\qed\enddemo
\demo{Proof of Theorem 3.1} Each fundamental solution $\psi_{\nu}=\mn e^\Phi$
differs from the solution constructed in Lemma 3.2 by right multiplication
by an invertible constant matrix $C_{\nu}$.\qed\enddemo

Generically one has more information about the fundamental solution at
$z=0$ than is given above.  We begin with a general remark about the
constant term $A_0(x)=xq(x)+F_n(x)$.

\proclaim{\bf  Lemma 3.3} The isomonodromy equation (1.6) implies that the
matrices $xq+F_n$ are similar for all values of $x$.\endproclaim

\demo{Proof}  Let $a(x)$ be a non-singular solution of $da/dx = qa$.
Then
$$
\frac{d}{dx}\biggl[a^{-1}(xq+F_n)a\biggr] = a^{-1}
\biggl[\,\frac{d}{dx} - q\,,\,xq+F_n\,\biggr]a=0\,
$$
so $a\inv (xq+F_n)a$ is constant.\qed\enddemo

\proclaim{\bf  Corollary 3.4}  If $q$ and its derivatives of order less than $n$
have limit zero anywhere, then $xq+F_n$ vanishes identically.\endproclaim

\noindent{\bf Definition.} The matrix function $q$ is {\it proper} 
if $D_z\psi=0$ has a fundamental solution  of the form
$$
\psi(x_0,z)=w(x_0,z)z^{A_0(x_0)},\quad w(x_0,\cdot\,)\ \ \text{entire,}
\quad w(x_0,0)=\pmb 1.\tag3.3
$$
for some $x_0\in I$.  
\ss
\noindent{\bf Remarks.} This is a condition on the $(n-1)$-jet of $q$ at $x_0$.
By the proof of Lemma 3.2 
the condition is in fact independent of $x_0$, and applies to all $x\in I$.

In particular, $q$ is proper if for some $x_0$ no two eigenvalues of $A(x_0)$
differ by a non-zero integer (CL, Theorem 4.4.1]); hence
generically, $q$ is proper. 
\ss
When the eigenvalues of $A_0(x_0)$ differ by an integer,
 there is still a solution at
the origin of the form $wz^B$, with $w$ entire ( [CL], Ch. 4, Theorem 4.2);
but generically, $w$ is degenerate at the origin, and $B$ need 
no longer be similar to $A_0(x_0)$. ($B$ is similar to $A_0(x_0)$
if $w$ is invertible at the origin.)

These two situations are simply illustrated by the Painlev\'e II equation.
\ss
\noindent{\bf Example:} {\sl Half-Integer Solutions of Painlev\'e II.} 
As in \S 1, we take
$$
\gather
J=\mu=\lmat 1 & 0 \\ 0 & -1 \rmat, \qquad q=u\sigma=u
\lmat 0 & 1 \\ 1 & 0 \rmat; \\
F_1=q, \qquad F_2=\frac{1}{2}\lmat -u^2 & u_x \\ -u_x & u^2 \rmat,
\qquad F_3=\frac14\lmat 0&u_{xx}-2u^3\\u_{xx}-2u^3 & 0\rmat.
\endgather
$$
As noted earlier the isomonodromy equation is
$$
\tfrac{1}{4}u_{xx}-\tfrac{1}{2}u^3+xu=\nu \tag PII
$$
where $\nu$ is a constant; and $A_0(x)\equiv \nu \sigma$ is constant.

Let us consider the case $\nu=1/2$ and look for a solution of the
form $wz^{\sigma /2}$. We pick an initial point $x_0$ and let 
$$
u(x_0)=\a\,, \qquad u^{\prime}(x_0)=\b\,.
$$
The equation $D_z(wz^{\sigma /2})=0$ is
$$
z\frac{d}{dz}( wz^{\sigma /2})=zw_z z^{\sigma/2}+wz^{\sigma/2}\sigma
=(A_0+A_1z+\dots)wz^{\sigma/2}.
$$
Substituting a power series expansion $w=w_0+w_1z+\dots$
into this equation, we get the recursion
relations
$$
[w_0,\sigma]=0, \qquad w_1+\frac{1}{2}[w_1,\sigma]=A_1w_0,
\qquad \dots,
$$
where
$$
A_1=F_2+xq=\lmat x_0-\a^2/2 & \b /2 \\ -\b /2 & -x_0+\a^2 /2 \rmat.
$$
From the first equation, $w_0$ commutes with $\sigma$; and since the equations are invariant under
right multiplication by a constant matrix, we may factor out $w_0$ 
on the right.
Hence without loss of generality we may take $w_0=\pmb 1$. Taking
$$
w_1=\lmat a & b \\ c & d \rmat
$$
and turning to the second recursion relations, we find $a+d=b+c=0$ 
and the constraint
$$
\b+\a^2=2x_0. \tag C
$$ 
Since this constraint
is independent of the initial point $x_0$ we get the additional
equation
$$
u^{\prime}+u^2=2x.
$$
It is easily seen that any solution of this equation is also a
solution of (PII). Moreover, setting $u=\chi'/\chi$
one finds that $\chi$ satisfies the Airy equation
$$
\chi''-2x\chi=0.
$$
These special solutions were originally discovered by Gambier [Ga];
other special half-integer solutions were constructed by
Airault [Ai], using B\"acklund transformations.

The constraint $C$ on the 1-jet of the solution at $x_0$ shows that
the proper solutions of Painlev\'e II for $\nu=1/2$ form a 
submanifold of codimension 1, and so are non-generic.
The same considerations apply in general to solutions of the isomonodromy equations
when the eigenvalues of $A_0$ differ by integers. The monodromy data
for the general case
is discussed in Theorem 4.1.

It is important to note that $q$ {\it need not} be proper.  The condition
that it be proper is a non-trivial condition on the $(n-1)$--jet of
$q$, which determines the operator $D_z$; see Chapter 4 of [CL].

\proclaim{\bf  Theorem 3.5}  Suppose that $q$ is proper.  Then 
 there are unique constant matrices $\Bn$ 
such that the functions $\pn(x,z)z^{-\Bn}$ are entire functions of 
$z$, invertible at $z=0$.  Each $\Bn$ is similar to $A_0(x_0)$
and therefore has trace $0$.  Moreover
$$
B_{\nu+1}=\Sn^{-1}\Bn\Sn\,;\qquad \exp(2\pi i\Bn)= \Sn S_{\nu+1}\cdots
S_{\nu-1}.\tag3.4
$$\endproclaim
\demo{Proof} Combining the assumption that $q$ is proper with the 
construction in Lemma 3.3, we obtain a fundamental solution 
$$
\psi(x,z) = w(x,z)z^{A_0(x_0)} 
$$
with $w(x,.)$ entire and $w(x_0,0)=\pmb 1$. Then 
there are constant matrices $\Cn$ such that 
$$
\pn = \psi\Cn = w\Cn z^{\Bn}\,,\qquad \Bn = \Cn^{-1}A_0(x_0)\Cn.\tag3.5
$$
The first relation in (3.4) follows from (2.4).  The second relation
follows from the fact that the analytic continuations of $\pn$ and
$z^{\Bn}$ around the origin are $\pn\Sn S_{\nu+1}\cdots S_{\nu-1}$
and $z^{\Bn}\exp(2\pi i\Bn)$ respectively. 

If $wz^B = \tilde w z^{\tilde B}$
with $w$ and $\tilde w$ regular and invertible at the origin, then
$z^{\tilde B}z^{-B}$ is regular at the origin and it follows that $\tilde
B = B$.  Thus the $\Bn$ are unique.
\newpage

\bf 4. The inverse problem and the Painlev\'e property\rm
\vskip .3cm

The inverse problem, to determine the solution of the isomonodromy
equation from its monodromy data, can be formulated as a Riemann-Hilbert
problem. This idea was introduced in the context of evolution equations by
Shabat and utilized by Flaschka and Newell in $2\times 2$ cases.
We give the general formulation in this section.  We also show that
{\it every} solution of a Riemann-Hilbert problem of this type is
associated to a solution of an isomonodromy equation (1.6).

It has been shown by Malgrange [Ma] and Miwa [Mi] that equations like
those considered here have the Painlev\'e property.  We sketch here an
argument based on the Riemann-Hilbert problem.
 
Suppose again that $q$ is a solution of the isomonodromy equation.

\proclaim{\bf  Theorem 4.1} The function $q$ is uniquely determined by its
Stokes matrices $\Sn$, the function $\Phi(x,z)=\frac1n z^n\mu+xzJ$,
 and the functions $\Un$ of Theorem 3.1.

If $q$ is proper, then $q$ is uniquely determined by the Stokes
matrices $\Sn$, the function $\Phi$, and the exponents $\Bn$ of Theorem 3.5.
\endproclaim

\demo{Proof}  Let the functions $\pn$, $\mn$ be as defined previously.
Let $\Sigma_{\nu}$ be a ray in the intersection $\On\cap\Omega_{\nu+1}$,
let $\Gamma=\{|z|=1\}$ be the unit circle, and let 
$$
K = \Gamma\cup\biggl(\bigcup_{\nu}\Sigma_{\nu}\cap\{z:|z|>1\}\biggr)\,.
$$
Fix  $x\in I$ and define a function $M$ on $\Bbb C\setminus K$ by
$$\align
M(z)&=\pn(x,z)\Un(z)^{-1}\,,\quad\text{for $|z|<1$}\,\tag4.1\\
M(z)&=\mn(x,z)\quad\text{for $|z|>1$,\  $z$ lying between $\Sigma_{\nu-1}$
and $\Sigma_{\nu}$}\,.\tag4.2\endalign
$$
(Recall that by Theorem 3.1 the value given by (4.1) is independent of $\nu$.)
The matrix function $M$ has the properties: 
$$\align
M&\ \ \text{is holomorphic and invertible on}\ \Bbb C\setminus K\,;\tag4.3\\
M&\ \ \text{has limit $\pmb 1$ as $z\to\infty$}\,;\tag4.4\\
M&\ \ \text{is continuous up to the boundary from each component}\,.\tag4.5\\
\endalign
$$ 
Let $M_\Gamma$ and $\Mn$ denote the boundary values of $M$ on the circle from 
the unit disc and on the boundary of the region outside the circle between the
rays $\Sigma_{\nu-1}$ and $\Sigma_{\nu}$, respectively.
These boundary values are linked on their common domains of definition by
$$
\Mn(z) = M_{\Gamma}(z)\Un e^{-\Phi(x,z)}\,;\quad M_{\nu+1}(z)=\Mn e^{\Phi
(x,z)}\Sn e^{-\Phi(x,z)}\,.\tag4.6
$$
Uniqueness in the general case will follow if we show that $M$ is uniquely
determined by the Riemann-Hilbert problem (4.3)--(4.6).  But this is
standard: 
if $\tilde M$ is a second solution of (4.3)-(4.6), then $\tilde M M^{-1}$
is piecewise holomorphic and continuous, with value $\pmb 1$ at $\infty$,
so $\tilde M M^{-1}\equiv\pmb 1$.

Suppose now that $q$ is proper.  Then we redefine $M$ on the unit disc
by
$$
M(z)=\pn(x,z)z^{-\Bn}\ \ \text{if}\ \ |z|<1\,.\tag4.7
$$
Then $M$ has the previous properties, with (4.6) modified to
$$
\Mn(z) = M_{\Gamma}(z)z^{\Bn} e^{-\Phi(x,z)}\,;\quad M_{\nu+1}(z)=\Mn e^{\Phi
(x,z)}\Sn e^{-\Phi(x,z)}\,.\tag4.8
$$
As before, this $M$ is uniquely determined by the conditions (4.3)-(4.5)
and the boundary relations (4.8).
\qed\enddemo
Our principal result concerning the isomonodromy equation is the following.

\proclaim{\bf  Theorem 4.2} Any solution of the isomonodromy equation (1.6) 
on an interval or connected domain
has a single-valued meromorphic extension to all $x\in \Bbb C$.
In particular (1.6) has the Painlev\'e property: the only movable
singularities of its solutions are poles.
\endproclaim

\demo{Proof}  The Riemann-Hilbert problem (4.3) to (4.6) has data which
depend holomorphically on $x$ through the function $\Phi(x,z)
=\Phi(0,z)+xzJ$.

Since the ray $\Sigma_{\nu}$ lies in the sector $\On$, 
the off-diagonal elements of $e^{\Phi}\Sn e^{-\Phi}$ decay at infinity
like $\exp(-\epsilon |z|^n)$, which more than offsets any growth from
$\exp(xzJ)$ and its derivatives with respect to $x$.  

The modified Riemann-Hilbert problem can be written as a singular integral
equation with parameter $x$, which we write symbolically as 
$$
M = \pmb 1 + C_{x}\,M\,,\tag4.9
$$
where $C_x$ also depends on the data $S$, $U$.
The operator $Id-C_x$ depends holomorphically on $x$ in the entire plane 
and is invertible for all $x$ in the original domain $I$ of $q$.  
It is shown in [BC] that $Id-C_x$ is Fredholm with index zero; in fact
it is shown that inverting $Id-C_x$ can be accomplished (if at all) in
two steps: inverting a small perturbation of the identity, followed by
solving a finite system of linear equations which depend 
on $x$.  In the present case both the perturbation of the identity and
the system of linear equations depend holomorphically on $x$, and it
follows that the solution is meromorphic with respect to $x$ in the
entire $x$--plane.  Therefore the associated potential $q$ is also
meromorphic in the entire $x$--plane. (In [BC] the sufficient conditions
include compatibility conditions for the data at the intersections points
of its domain of definition; these conditions are shown in [BDT] to be
a consequence of a product condition on Taylor expansions of the data
at the intersection points.  In the present case the data is assumed to
have come from the direct problem, so
the conditions are automatically satisfied.)
\qed\enddemo

The next result closes the circle, insofar as proper solutions are 
concerned.  Recall the following facts about the Stokes matrices and
the matrices $\Bn$:
$$\gather
e^\Phi\Sn e^{-\Phi} \quad \text{is bounded as}\ \ z\to\infty,\ \  z\in\On\,,
\qquad(\Sn)_{jj} = 1\,;\tag4.10\\
\tr(\Bn)=0\,\qquad
\exp(2\pi i\Bn) = \Sn S_{\nu+1}\cdots S_{\nu-1}\,.\tag4.11\endgather
$$
\proclaim{\bf  Theorem 4.3} Suppose matrices $\Sn$, $\Bn$ satisfy the conditions
(4.10), (4.11).  If the associated Riemann-Hilbert problem (4.3)-(4.5), (4.8)
has a solution $M(x_0,z)$ for some value $x_0$ of $x$, then it has a solution 
$M(x,z)$ except
for a discrete set of values of $x\in\Bbb C$.  For $|z|>1$, $z\notin\bigcup
\Sigma_{\nu}$, the function $\psi(x,z)=M(x,z)e^\Phi$ satisfies the system of
equations
$$
D_x\psi = 0\,,\qquad D_z\psi = 0\tag4.12
$$
where the operators $D_x = \partial/\partial x - zJ - q$ and
$D_z = z\partial/\partial z - [z^nF]_+ -xzJ-xq$ are defined by a 
unique off-diagonal matrix function $q$.  The function $q$ satisfies
the isomonodromy equation and is proper. 
\endproclaim
\demo{Proof} Because of the properties of the Riemann-Hilbert problem
as indicated in the previous proof, solvability at $x=x_0$ implies
solvability except in a discrete subset of $\Bbb C$.  The determinant
satisfies a Riemann-Hilbert problem which forces $\det M\equiv 1$;
in fact since $\tr(\Phi)=\tr(\Bn)=0$, it follows that $\det M$ is
continuous across $K$ and hence entire in $z$, and (4.4) implies that
$\det M$ is $1$ at infinity.
Thus $M$ is invertible.  The function 
$$
\left[\left(\px-z\ad J\right)M\right]M^{-1}
$$
is piecewise holomorphic, continuous, and bounded, hence has value
$q(x)$ independent of $z$.  Therefore $\psi=Me^\Phi$ satisfies
the first of the equations (4.12), for this choice of $q$. 
Moreover $M(x,\cdot)$ has an asymptotic expansion $\sum z^{-j}f_j(x)$
as $z\to\infty$ and $q = -[J,f_1]$ is off-diagonal.
On the other hand, the function
$$
\left[\left(\pz-z^n\ad\mu-xz\ad J\right)M\right]M^{-1}
$$
is piecewise holomorphic, continuous, and $O(z^{n-1})$ as $z\to\infty$,
hence is a polynomial of degree less than $n$ in $z$.  Therefore $\psi$
satisfies an equation
$$
\pz\psi=A(x,z)\psi= \sum^n_{j=0}z^jA_j(x)\,,\quad\ A_n = \mu\,.
$$
As in the proof of Theorem 1.1, we conclude from the asymptotic
 expansion of $M$ at infinity that
$$
A(x,z)= [z^nF]_+ + xzJ + xq\,.
$$
Thus $\psi$ satisfies the equations $D_x\psi=0$ and $D_z\psi=0$, 
so $q$ satisfies the isomonodromy equation.  

Finally, we must show that $q$ is proper.  Let $\pn$ be the restriction
of $\psi$ to $\On\cap\{|z|>1\}$ and extend $\pn$ to the unit disc by
setting $\pn = Mz^{\Bn}$ for $|z|<1$.  Then $\pn$ is continuous across
the circular arc, because of (4.8),
and is therefore holomorphic.  Since $M$ is regular at
the origin, $\pn$ has the desired form and $q$ is proper.
\qed\enddemo
\ms
{\bf 5. Rational solutions}
\ss
In this section we investigate rational solutions of the isomonodromy
equations.  These are the analogues of the reflectionless potentials
in inverse scattering theory; indeed
there is a close connection between the Stokes matrices
$\Sn$ being trivial ($\equiv \pmb 1$), the normalized wave functions $\mn$
being rational in $z$, and $q$ being rational in $x$.

\proclaim{\bf Theorem 5.1} Suppose that $q$ satisfies the isomonodromy
equation (1.6).  The associated system (1.8), (1.9) has a solution
$m$ which is rational in $z$ and equals $\pmb 1$ at $z=\infty$ if and only
if each Stokes matrix $\Sn$ equals \ $\pmb 1$. 

If also $q$ is proper, then $q$ and $m$ are rational functions of $x$.
\endproclaim
\demo{Proof}  If (1.8), (1.9) has a solution $m$ which is rational in
$z$ and equals $\pmb 1$ at $\infty$, then by uniqueness $\mn=m$, all $\nu$,
and so each $\Sn=\pmb 1$.  Conversely, suppose each $\Sn=\pmb 1$.  Then
the $\mn$ coincide and so $m=\mn$ is regular on the Riemann sphere minus
the origin.  It follows that in the representation (3.1), 
$m=f(x,z)V(z)z^B$ with $fV$ entire,  
the factor $z^B$ is single-valued.  Therefore $\exp(2\pi i
B)=\pmb 1$, and therefore $B$ is diagonalizable, with integer eigenvalues.
This shows that the origin is a pole for $m$, hence $m$ is rational.

Now suppose that $q$ is proper, i.e. that $B$ has trace zero.
Suppose that $m$ has a pole of order $r$
at $z=0$.  Then the smallest eigenvalue of $B$ is $-r$.  Denote the 
dimension of the eigenspace corresponding to the eigenvalue $s$ by
$d_s$.   Let $(v_1,v_2,\dots,v_d)$ be a basis of
eigenvectors of $B$ with eigenvalues $s=\a_1\geq\a_2\geq\dots\geq\a_d=
-r$.  For $1-r\leq k\leq s=\a_1$ define a $d\times (d_k+\dots d_s)$
matrix $V_k$ by taking the columns to be those $v_j$ with $\a_j\geq k$.
Let 
$$
e^{\Phi(x,z)}=\sum^\infty_{j=0} z^jC_j(x),
$$
where the matrix-valued function $C_j$ is a polynomial of degree $j$ in
$x$.  Then the condition  
$$
m\,e^\Phi z^{-B}=
[\pmb 1+z\inv f_1+\dots +z^{-r}f_r]\,e^\Phi z^{-B}\quad\text{is regular
at}\ \ z=0\tag5.1
$$
is equivalent to the sequence of equations
$$\align
f_r\, V_{1-r}&=0,\tag5.2\\
[f_r C_1+f_{r-1}]\,V_{2-r}&=0,\\
& \dots\\
[f_r C_{r-1}+\dots+f_1]\,V_0&=0,\\
[f_r C_r+\dots+f_1C_1]\,V_1&=-V_1,\\
&\dots\\
[f_r C_{s+r-1}+\dots+ f_1C_s]\,V_s&=-V_s.
\endalign
						$$
The first equation in (5.2) represents $d(d_{1-r}+d_{2-r}+\dots\ )$
linear equations for the $d^2$ entries of $f_r$.  Altogether the number of 
linear equations represented by (5.2) is
$$\align
d\,[(d_{1-r}+&d_{2-r}+\dots\ )+(d_{2-r}+d_{3-r}+\dots\ )+(d_{3-r}+\dots )+\dots
\ ]\\
&= d\,[d_{1-r}+2d_{2-r}+3d_{3-r}+\dots ]\\
&= d\,[(1-r)d_{1-r}+(2-r)d_{2-r}+\dots]+
d\,r\,[d_{1-r}+d_{2-1}+\dots\ ]\\ 
&=d\,[\tr(B)+rd_{-r}]+d\,r\,[d-d_{-r}]=rd^2.\endalign
$$
Thus (5.2) is a set of $rd^2$ equations for the $rd^2$ entries of the $f_j$.
As noted above, the coefficients in these equations are polynomials in $x$.
To prove $m$ is rational in $x$ we only need to show that the system has a 
unique solution. By assumption it has one solution.  As noted, (5.2) 
is equivalent to (5.1).  Note also that
$$
\lim_{z\to\infty}\det(fe^\Phi z^{-B}) = 1\,,
$$
so (5.1) implies that $fe^\Phi z^{-B}$ is invertible for all $z$.
Consequently, if $\tilde f$ is a second solution of (5.1) then
$\tilde f\,f\inv$ is rational, entire, and $\pmb 1$ at $\infty$,
hence identically $\pmb 1$.
Thus the solution is unique and $m$ is rational in $x$.  Now the
potential $q=-[\,J\,,\,f_1\,]$, so $q$ is also rational.
\enddemo

Conversely, the equations (5.2) allow construction of rational solutions
of the isomonodromy equation. 

\proclaim{\bf Theorem 5.2}  Suppose that $B$ is a constant matrix 
with trace zero and minimum eigenvalue $-r$, and that
$\exp(2\pi iB)=\pmb 1$.  Suppose that 
the equations (5.2) have a solution.  Then $q=-[\,J\,,\,f_1\,]$ is
rational and is a proper solution of the isomonodromy equation 
having normalized wave function $m=\pmb 1+z\inv f_1+\dots+z^{-r} f_r$.
Both $m$ and $q$ are rational functions of $x$.
\endproclaim

\demo{Proof} In view of the previous proof, all we need to show is
that $m=\pmb 1+z\inv f_1+\dots+z^{-r} f_r$ satisfies equations of the
form (1.8), (1.9).  Let $\psi=me^\Phi=wz^B$.
By assumption, $w$ is entire and invertible everywhere.  Consider
the function
$$
\biggl[\biggl(\dx-zJ\biggr)\,\psi\biggr]\,\psi\inv
=\pd{w}{x}\,w\inv-zJ\,.
$$
which is entire.  The behavior at $z=\infty$ is obtained by rewriting
the function as
$$
\biggl[\biggl(\frac{d}{dx}-z\ad J\biggr)\,m\biggr]\,m\inv
=-[\,J\,,\,f_1\,]+O(z\inv)\,.
$$
Thus by Liouville's theorem  the function is independent of $z$ and
is identically $q(x)=-[\,J\,,\,f_1\,]$, and $m$ satisfies (1.8) with
this choice of $q$.  Similarly, consider the function
$$
A(x,z)=\biggl[z\pd{\psi}{z}\biggr]\,\psi\inv
=\biggl[\pz\biggl(wz^B\biggr)\biggr]\,z^{-B}w\inv
=\biggl[z\pd{w}{z}\biggr]\,w\inv+B\,.
$$
It is regular at $z=0$ and its behavior at $z=\infty$ is seen
by rewriting it in terms of $m$:
$$
\biggl[\pz\biggl(me^\Phi\biggr)\biggr]\,e^{-\Phi}m\inv
=\biggl[z\pd{m}{z}\biggr]\,m\inv+m\,(z^n\mu+xzJ)\,m\inv
=z^n\mu+O(z^{n-1})\,.
$$
Thus $A$ is a polynomial in $z$ with leading term $z^n\mu$, and 
$m$ satisfies (1.9).  This completes the proof.
\enddemo

We turn now to a closer inspection of the system (5.2), particularly
the behavior for large $x$.  
Note that the coefficients $C_j$ of the expansion of $e^\Phi$
have the form
$$
C_j(x)=\frac1{j!}x^jJ^j+O(x^{j-n}).\tag5.3
$$
Rescale the problem for large $x$ by setting $x^jf_j=\tilde f_j$.
Then in view of (5.3) the equations (5.2) have the form
$$\align
\tilde f_r V_{1-r}&=O(x^{-n}),\tag5.4\\
[\tilde f_r J +\tilde f_{r-1} ]\,V_{2-r}&=0(x^{-n}),\\
&\dots\\
[\tfrac1{(r-1)!}\tilde f_r J^{r-1}+\dots+\tilde f_1]\,V_0&=0(x^{-n}),\\
[\tfrac1{r!}\tilde f_r J^r+\dots+\tilde f_1 J]\,V_1&=-V_1+0(x^{-n}),\\
&\dots\\
[\tfrac1{(s+r-1)!}\tilde f_rJ^{s+r-1}+\dots+\tfrac1{s!}\tilde f_1J^s]\,V_s&=
-V_s+0(x^{-n}).
\endalign
$$
The key to the behavior for large $x$ is thus the $rd\times rd$ matrix
$$
\lmat V_{1-r}&JV_{2-r}&\dots& \tfrac1{(r-1)!} J^{r-1}V_0&\dots&
\tfrac1{(s+r-1)!}J^{s+r-1}V_s\\
0 &V_{2-r}&\dots&\tfrac1{(r-2)!}J^{r-2}V_0&\dots &\tfrac1{(s+r-2)!}
J^{s+r-2}V_s\\
         &&\ddots&&&\\
  0 &\dots&&V_0&\dots& \tfrac1{s!}J^sV_s\rmat.\tag5.6
$$

Since the $V_j$ are constructed from eigenvectors of $B$, 
invertibility of (5.6) is a
condition on the matrices $J$, $B$ alone.

\proclaim{\bf Theorem 5.3}  Suppose that $e^{2\pi i B}=\pmb 1$, the least
eigenvalue of $B$ is $-r$, and 
the matrices $J$ and $B$ are such that 
the matrix (5.6) is invertible. 
Then the equations (5.2) are solvable for all large $x$ and
the associated potential $q$ is rational; moreover
$q(x)=x\inv q_0+O(x^{-1-n})$, where the constant matrix $q_0$ is similar
to $B$.
\endproclaim

\demo{Proof} The preceding discussion establishes that invertibility
of (5.6) implies
solvability of (5.2) for large $x$.  It is clear from (5.6) that the
$f_j$ are of the form $x^{-j}\tilde f_j =x^{-j} f_{j0}+O(x^{-j-n})$ 
with $f_{j0}$ constant.
Therefore $q=-[\,J\,,\,f_1\,]$ has the form asserted above.  Finally,
it follows by induction that $F_j$ of Theorem 1.1 is $O(x^{-j})$.
Now $B$ is similar to $F_n+xq=q_0+O(x^{-n})$ for every $x$, so
$B\sim q_0$.\enddemo 

\ss
\flushpar{\bf Example.} As a simple example to illustrate
the construction of rational solutions we consider the AKNS
hierarchy; cf. also [FN].  Here
$$
J=\mu=\sigma_3=\lmat 1& 0\\0& 1\rmat, \qquad q=\lmat 0 & u \\ v & 0 \rmat.
$$
From the recursion relations (1.5) we find
$$
F_0=\sigma_3,\qquad F_1=q, \qquad F_2=\frac12\lmat -uv & u_x \\
-v_x & uv \rmat
$$
and the isomonodromy deformation equations (1.6) at order 2 are
$$
(xu)_x+\tfrac{1}{2}u_{xx}-u^2v=0, \qquad (xv)_x+\tfrac{1}{2}v_{xx}+uv^2=0.
$$
Under the symmetry reduction $u=\pm v$ these equations reduce to the 
single equation
$$
(xu)_x+\tfrac{1}{2}u_{xx}\pm u^3=0.
$$

For the case $r=1$ equations (5.3) are simply
$$
f_1v_1=0, \qquad (\pmb 1+f_1C_1)v_1=0,
$$
where $C_1=x\sigma_3$, and $v_1$ is the eigenvector
of $B$ with eigenvalue 1. Since we do not know $B$ {\it a priori},
let us take $v=[1,\ h]^t$
where $h$ is to be determined. We find
$$
f_1=\frac{1}{2x}\lmat -1 & h^{-1} \\ -h & 1 \rmat,
\qquad m=\pmb 1+z\inv f_1, \qquad
q=-[\sigma_3,f_1]=-\frac{1}{x}\lmat 0 & h^{-1} \\ h & 0 \rmat.
$$
It is easily verified that $\det\, ( \pmb 1+f_1z^{-1})=1$
for all  $h$. Moreover
this result is independent of $n$, the order of the equation in
the hierarchy. Thus for all $h$ we get a solution with
$$
u=-\frac{1}{hx}, \qquad v=-\frac{h}{x}, \quad h\ne 0.\tag5.8
$$

The construction of rational solutions for $r=2$ is somewhat more
involved. Under the symmetry reduction $u=v$, 
we have $\sigma m(x,z)\sigma^{-1}=m(x,-z)$, where
$$
\sigma=\lmat 0 & 1 \\ 1 & 0 \rmat;
$$
hence
$$
f_j=\lmat a_j & (-1)^jb_j\\ b_j & (-1)^j a_j \rmat.
$$
For $n=2$, 
$$
C_1=x\sigma_3, \qquad C_2=\frac{x^2}{2}\pmb 1 +\sigma_3,
\qquad C_3=\frac{x^3}{3!}\sigma_3+x\pmb 1.
$$
In the case $r=2$, and $h=1$ one finds
$$
q=-[\sigma_3,f_1]= \biggl(\frac{3(x^2+1)(x-1)}{x(x^4+3)}-\frac{1}{x}
\biggr)\sigma.
$$

It is also interesting to consider the symmetry reduction
$v=\epsilon \bar u$ where $\epsilon=\pm 1$. Replacing $t$
by $it$ in the time evolution equation
$q_t=[\px - q, F_2]$, one obtains the nonlinear Schr\"odinger equation
$$
iu_t=\frac{1}{2}u_{xx}-\epsilon |u|^2u. \tag5.9
$$
There are solitons only in the case $\epsilon=-1$, called the 
self-focussing case. This equation is invariant under the scaling
\def\tt{\rightarrow}
\def\l{\lambda}
$u(x,t)\tt \l u(\l x,\l^2 t)$
and the associated similarity solutions take the form
$$
u(x,t)=\frac{1}{\sqrt t}\phi (\xi), \qquad \xi= \frac{x}{\sqrt t}
$$
where $\phi$ satisfies the ordinary differential equation
$$
i(\xi \phi)^{\prime}+2\epsilon |\phi|^2\phi -\phi^{\prime \prime}=0.
\tag5.10
$$

We can construct ``rational solutions" of this equation for
real $x$, although they cannot be continued into the
complex $x$ plane. For example, taking $h=i$ in (5.8) we obtain
$$
u=\frac{i}{x}, \qquad v=\frac{-i}{x}=\bar u
$$
for real $x$. Hence we obtain rational solutions of (5.10) for real
$x$ in the defocussing case. 

In the self-focussing case, $xq+F_2$ is skew-Hermitian, so always
has imaginary eigenvalues. Therefore the solutions
$wz^{A_0}$ have non-trivial monodromy in a neighborhood of the origin,
and solutions of (1.8), (1.9) for $m$ rational in $z$ do not exist.
Thus the theory of the present section does not apply.

\ms
{\bf 6. B\"acklund transformations}
\vskip .3cm
In this section we adapt a procedure which goes back at least to Moutard
[Mt] for transforming a pair consisting of a linear differential operator
and its eigenfunctions to a new such pair.  Here the idea is to make a
gauge transformation of normalized solutions $\psi$ associated to a given
problem.  We look for a gauge transformation of the form
$$
\tilde\psi(x,z)=G(x,z)\psi(x,z)=\bigl[\pmb 1-z\inv a(x)
\bigr]\psi(x,z)\,.\tag6.1
$$
We assume that $\psi$ is invertible and 
satisfies (1.1), (1.2), where $q$ satisfies the isomonodromy equation.
Then $\tilde\psi$ satisfies
$$\align
\pd{\tilde\psi}{x}&=[G(Jz+q)G\inv-G_xG\inv]\,\tilde\psi\,,\tag6.2\\
z\pd{\tilde\psi}{z}&=[GA(z,x)G\inv-zG_zG\inv]\,\tilde\psi=
\tilde A(x,z)\,\tilde\psi\,.\tag6.3\endalign
$$
In order for 6.2 to take the form $\partial\tilde\psi/\partial x=
(Jz+\tilde q)\,\tilde\psi$ it is necessary and sufficient that
$$
\tilde q = q+[\,J\,,\,a\,]\,,\qquad \frac{da}{dx} =[\,q+Ja\,,\,a\,]\,.\tag6.4
$$
This matrix Riccati equation for $a$ can be linearized by setting
$a=bcb\inv$
 with $c$ constant; then the equation for $a$ is satisfied provided
$db/dx=qb+Jbc$.  In particular, there is a solution for each choice
of $c$  and each choice of (invertible) initial value $b(x_0)$.
Note that $c^2=0$ implies $a^2\equiv 0$.

The next problem is to ensure that $\tilde A(x,\cdot)$ in (6.3) is a
polynomial with leading term $z^n\mu$.  

\proclaim{\bf Theorem 6.1} Suppose $a$ satisfies (6.4), and $a^2\equiv 0$.
Suppose that $\tilde A(x_0,\cdot)$ is a polynomial in $z$.  Then
$\tilde A(x,\cdot)$ is a polynomial in $z$ for each $x$ for which $A$ is
defined, and $\tilde q$ satisfies the isomonodromy equation.\endproclaim
\demo{Proof}  The assumption $a^2=0$ implies that $G\inv=\pmb 1+z\inv a$,
so
$$
\tilde A(x,z)=\bigl[\pmb 1 -z\inv a(x)\bigr]A(x,z)
\bigl[\pmb 1+z\inv a(x)\bigr] - z\inv  a(x)\,.\tag6.5
$$
The zero-curvature condition (1.3) is preserved by gauge transformations, so
$$
\bigl[\,\px - zJ-\tilde q\,,\,\tilde A\,\bigr] = zJ\,.\tag6.6
$$
Equation (6.5) shows that $\tilde A$ has highest term $z^n\mu$ and
(6.6) shows that if $A(x,\cdot)$ is regular at $z=0$ for one value
of $x$, then this is true for all values of $x$.
To show that (6.6) amounts to the isomonodromy equation for $\tilde q$,
we must relate $\tilde A$ to the coefficients of the expansion of
$\tilde m\mu\tilde m\inv$, where $\tilde m =\tilde\psi e^\Phi$
The required relation (1.4) follows from the second part of Theorem 1.1:
let $\pn =\mn e^\Phi$ be the
normalized wave function for the sector $\On$.   Then $\tilde\mn=G\mn$
has an asymptotic expansion as $z\to\infty$ in $\On$ with leading term
 $\pmb 1$, as required.  Thus $\tilde q$ satisfies the isomonodromy
equation.
\enddemo

We now consider the case of proper $q$, i.e. $q$ for which the
fundamental solutions have the form (3.3) at $z=0$.

\proclaim{\bf Theorem 6.2} Suppose that $q$ is proper and that $xq+F_n$
is diagonalisable.  Then either $q$ satisfies an algebraic differential
equation of degree $n-1$, or there is a 
nontrivial B\"acklund transformation $q\rightarrow
\tilde q$ of the form (6.4) such that the eigenfunctions transform by
(6.1) and such that $\tilde q$ satisfies the isomonodromy equation.
Moreover, $\tilde q$ is also proper and $x\tilde q +\tilde F_n$
is diagonalizable.\endproclaim
\demo{Proof} Fix $x_0$ and write matrices with respect to a basis
for which $A_0(x_0)$ is diagonal.  Choose a solution of the form
(3.3), so $w(x_0,0)=\pmb 1$.  Then $\det w\equiv 1$.
Write $w(x_0,z)=\sum z^k w_k$.
Suppose that for some $j\neq k$, $(w_1)_{kj}\neq 0$, and take
$$
a_{jk}w_{1,jk} = 1\,;
\qquad\text{all other entries of }\ a_{rs}(x_0) = 0.\tag6.7
$$
Then a simple calculation shows that the $jj$--entry of 
$G(x_0,z)w(x_0,z)$ vanishes at $z=0$, and
all entries except the $jk$--entry are regular at $z=0$.  Therefore
$$
G(x_0,z)w(x_0,z)=\tilde w(x_0,z)z^C\,,
$$
with $\tilde w$ entire, $\det\tilde w\equiv 1$, $C$ diagonal,
$C=e_{jj}-e_{kk}$.
It follows that 
$$
\tilde A(x_0,\cdot) = \bigl[\pz \tilde w + \tilde w B\bigr]\tilde
w\inv\,,\quad B=C+A_0(x_0)\,.
$$
Thus $\tilde A(x_0,\cdot)$ is regular at $z=0$. We can prolong
$a$ by (6.4) and deduce the result from the preceding theorem.  
Clearly $\tilde q$ is regular at $x_0$ and hence for all $x$. 

The obstruction to this construction at $x=x_0$ is the vanishing of 
the off-diagonal entries of $w_1(x_0)$, so $[\,A_0\,,\,w_1(x_0)\,]=0$.
The equation for $w$ is
$$
z\frac{dw}{dz}=A(z,w)w-wB\,.
$$
Substitution of the power series for $w$ into this equation yields
$w_1=[A_0,w_1]+A_1$; hence one would have $[A_0,A_1]=0$ at $x_0$.
Thus the obstruction is the vanishing
of this commutator at every point, which is the algebraic ODE for
$q$ of order $n-1$:
$$
[\,F_n+xq\,,\,F_{n-1}+xJ\,]=0\,.
$$ 
\enddemo

A consequence of the argument just given is that the matrix $a$
for the transformation in Theorem 6.2 can be constructed algebraically,
bypassing the differential equation in (6.4), by 
looking at $w(x,\cdot)$ in a basis which diagonalizes $A_0(x)$.
Note
that $x\tilde q + \tilde F_n$ is similar to $B$ and hence diagonalizable.
Relative to $xq+F_n$, one eigenvalue has been increased by $1$ and a
second eigenvalue has been decreased by $1$.

\noindent{\bf Example.} We illustrate the method to construct
integer solutions of the Painlev\'e II equation by B\"acklund transformations 
(cf. also Airault [Ai]); we use the notation in \S3.

The  B\"acklund transformation from the zero solution to the solution for $r=1$ coincides
with the reduced wave function
$m$. In fact, (1.8) is equivalent to the intertwining
relation 
$$
m\biggl( \px -zJ\biggr)=\biggl(\px -zJ-q\biggr)m.
$$
We work in a basis in which $q$ is diagonal. Note that
$$
P\inv \sigma P=\sigma_3, \qquad P\inv \sigma_3 P=-\sigma,
\qquad P=\frac{1}{\sqrt 2}\lmat 1 & -1 \\ 1 & 1 \rmat.
$$
Thus in this basis $J=-\sigma$.

We look for a gauge transformation of the form
(6.1) with $a(x)=v(x)T$. In order that $a$ satisfy
(6.4) we must have
$$
-[\,\sigma T\,,\,T]=\lambda\, T, \qquad T^2=0
$$
for some scalar $\lambda$. These equations are satisfied
by the choices $T=N_{\pm}$:
$$
N_+=\lmat 0 & 0 \\ 1 & 0 \rmat
\qquad
N_-=\lmat 0 & 1 \\ 0 & 0 \rmat.
$$
The Riccati equation (6.4) is
$$
v_xN_{\pm}=v^2[\,-\sigma N_{\pm}\,,\,N_{\pm}\,]=v^2N_{\pm},
$$
hence
$$
v_x=v^2 .\tag6.8
$$
We use (6.7) to get the initial condition at an arbitrary point $x_0$. 
Since we are gauging from the trivial solution,
$w=\exp\{-(xz+z^3/3)\sigma\}$. Therefore, $w_1(x_0)=-x_0\sigma$,
while $a(x_0)=v(x_0)N_{\pm}$. In either case, (6.7) implies
$$
v(x_0)=-\frac{1}{x_0}.
$$
Since $x_0$ was arbitrary (other than $x_0=0$)
we see that $v(x)\equiv-1/x$, and this function
satisfies (6.8). Thus, as remarked above, $v$
is fully determined by the
algebraic condition (6.7).

The corresponding potentials and wave functions are
$$
q_{\pm}=[J,vN_{\pm}]=\pm \frac{1}{x}\sigma_3,
\qquad
m_{\pm}=I+\frac{1}{xz}N_{\pm}.
$$

Let us take the case $q=q_+$ and construct the gauge 
transformation from $r=1$ to $r=2$. We begin by determining
$w_1$. The wave function for $q_+$ is
$$
m_+e^{\Phi}=\lmat 1 & 0 \\ (xz)\inv & 1 \rmat
e^{-(xz+z^3/3)\sigma}.
$$
Expanding the exponential to third order terms in $z$ we find 
$$
\align
m_+e^{\Phi}=&\left\{ \lmat 0 & -x \\ x\inv & 0 \rmat
+z\lmat 1 & 0 \\ 0 &w\rmat
+\dots \right\} \lmat z\inv & 0 \\ 0 & z \rmat \\
=&\left\{ \pmb 1 + z \lmat 0 & x \\ -w & 0 \rmat
+\dots \right\}\lmat z & 0 \\ 0 & z\inv \rmat
\lmat 0 & -x \\ x\inv & 0\rmat.
\endalign
$$
where
$$
w=\frac{x^3-1}{3x}.
$$
Therefore
$$
w_1=\lmat 0 & x \\ -w & 0 \rmat
$$
in (6.7). Then with $k=2,j=1$ we take
$$
v(x)=-\frac1w=\frac{3x^2}{1-x^3}.
$$
This choice indeed satisfies the Riccati equation (6.4), which in this
case is
$$
v_x=\frac{2v}{x}+v^2.
$$

The potential and wave function for $r=2$ is
$$
q=-\left(\frac{6x^2}{1-x^3}+\frac{2}{x}\right), \qquad
m=\lmat 1 & -\frac{3x^2}{(1-x^3)z} \\ 0 & 1 \rmat
\lmat 1 & 0 \\ (xz)\inv & 1 \rmat.
$$
The gauge transformation $G$ is the first factor of $m$.

\ms
{\bf 7. Scaling, self-similarity, and construction of isomonodromy
deformations}   
\ss
The isomonodromy deformations (1.1) and (1.2) can be derived 
from the Lax pairs for
integrable systems by a scaling invariance.  This was done for the 
mKdV equation by Flaschka and Newell [FN], working with the $2\times 2$
AKNS system.  For the general $d\times d$ system (1.1) and a given
constant matrix $\mu$, the $n$-th equation in the associated
hierarchy of equations is
$$
\pd{q}{t}=[\,J\,,\,F_{q,n+1}\,]=\biggl[\,\pd{}{x}-q\,,\,F_{q,n}\,
\biggr]\,.\tag7.1
$$
Here $F_{q,k}$ is the coefficient of $z^{-k}$ in the formal
series of Theorem 1.1.  
This equation is associated to the operator
$$
D_{t,q}=\pd{}{t}-[z^n F_q]_+\,,
$$
where $[z^nF_q]_+$ denotes the polynomial part of $z^n F_q$.
In fact (7.1) is equivalent to the zero-curvature condition
$$
[\,D_{x,q}\,,\,D_{t,q}\,]=0\,,\tag7.2
$$
which is the compatibility condition for the overdetermined system
$$
D_{x,q}\psi = 0\,,\qquad D_{t,q}\psi = 0\,,\tag7.3
$$
where 
$$
D_{x,q}=\pd{}{x}-zJ-q\,.
$$
A solution $q=q(x,t)$ of (7.1) is said to be {\it self-similar} if
$$ 
q(x,t)= q_\l(x,t)\equiv\l\inv q(\l\inv x, \l^{-n}\,t)\,,\qquad\l>0\,.\tag7.4
$$
Note that such a function is uniquely determined by its values at fixed $t$,
say at $t=1/n$.

Define dilations $\tl$, $\l>0$, acting on functions of $(x,t,z)$
by 
$$
\tl f(x,t,z)=f(\l x, \l^n\, t, \l\inv z)\,.\tag7.5
$$
Let $q$ be self-similar; then 
$$
D_{x,q}\,\tl = \l\tl\, D_{x,q}\,,\qquad D_{t,q}=\l^n\tl\,D_{t,q}\,.\tag7.6
$$
The first of these operator identities is immediate from (7.4).  For the
second, let $m_0(x,z)$ be a formal solution of $D_{x,q}m=-zmJ$ at $t=1/n$
and extend $m_0$ as a function of $t$ by
$m(x,t,z)=m_0((nt)^{-1/n}x, (nt)^{1/n}z)$.  
Then $m$ is invariant under the dilations,
so the first identity in (7.6) shows that $m$ is a formal solution
for each fixed $t$.  It follows that $F_q=m\mu m\inv$ is dilation
invariant,  which implies the second identity in (7.6).  A consequence
is that the evolution equations (7.1) are invariant under dilation.

The basic observation is that self-similar solutions of the
evolution equation (7.1) correspond to solutions of (1.6)
and conversely.

\proclaim{\bf Theorem 7.1} If $q(x,t)$ is a self-similar solution of
(7.1), then $q(x,1/n)$ is a solution of the isomonodromy equation
$$
\biggl[\dx-q\,,\,xq+F_n\,\biggr]=0\,.\tag7.7
$$
Conversely, if $q_0(x)$ is a solution of (7.7), then $q(x,t)
=(nt)^{-1/n}q_0((nt)^{-1/n}x)$ is a self-similar solution of (7.1)
with $q(x,1/n)=q_0(x)$.  
\endproclaim

\demo{Proof} Suppose $q$ is a self-similar solution of (7.1).
Differentiating (7.4) with respect to $\l$ at $\l=1$, we obtain
the Euler equation
$$
x\,\pd{q}{x}+nt\,\pd{q}{t}+q=0\,.\tag7.8
$$
At $t=1/n$ this identity becomes $q_t=-(xq)_x$.  Substituting this
last identity in (7.1), we obtain (7.7).

Conversely suppose that $q_0(x)$ satisfies (7.7).
Let $q$ be the self-similar extension.  By the argument leading to
the second identity in (7.6), $F_q$ is the invariant extension of $F$.
Consequently we only need to check (7.1)
or (7.2) when $t=1/n$, where it is implied by (7.7), (7.8).\enddemo

It is worth noting that the scaling argument carries over to the
wave functions.  Scaling invariance $\tl\psi=\psi$ is equivalent to
the condition that $\psi(x,1/n,z)$ satisfy (1.2).
In fact scaling invariance implies the Euler equation
$$
x\pd{\psi}{x}+nt\pd{\psi}{t}-z\pd{\psi}{z}=0\,,\tag7.10
$$
and, at $t=1/n$, (7.10) and (7.3) imply (1.1).  Conversely, (1.1) and
(7.10) for a function $\psi_0(x,z)$ imply that 
the self-similar extension $\psi(x,t,z)\equiv
\psi_0((nt)^{-1/n}x,(nt)^{1/n}z)$ satisfies (7.5).
\ss
Similar considerations apply to the Gel'fand-Dikii flows
$$
\dot L=[\,[L^{k/n}]_+ \,,\,L\,]\,,\qquad L=L_n=D^n+\sum^{n-2}_{j=0}
u_j(x,t)D^j\,,\qquad D=\frac{d}{dx}\,,\tag7.11
$$
where $k\in\Bbb N$ is not divisible by $n$ and $[L^{k/n}]_+$ denotes
the differential part of the pseudodifferential fractional power. 
The coefficients of $\L$ are polynomials in the $u_j$'s and their derivatives,
with no constant terms.
The simplest examples are
$$\align
L_2=D^2+u\,,\qquad&[L_2^{3/2}]_+=D^3+\tfrac32 u D+\tfrac34 u_x\,;\\
L_3=D^3+u_1 D + u_0\,,\qquad &[L_3^{2/3}]_+=D^2+\tfrac23 u_1\,,
\endalign
$$
These lead to the KdV and Boussinesq equations respectively:
$$
u_t=[\,[L_2^{3/2}]_+\,,\,L_2]=\tfrac14u_{xxx}+\tfrac32 uu_x\,;
$$
$$
u_{1t}D+u_{0t}=[\,[L_3^{2/3}]_+\,,\,L_3]=(2u_{0x}-u_{1xx})D+
(u_{0xx}-\tfrac23 u_{1xxx}-\tfrac23 u_1 u_{1x})\,.
$$
The next equation in the KdV hierarchy,
$$
u_t=\tfrac1{16}D^5u+\tfrac58 uu_{xxx}+\tfrac54u_xu_{xx}+\tfrac{15}8u^2u_x,
$$
comes from 
$$
[L_2^{5/2}]_+=D^5 + \tfrac52 uD^3+\tfrac{15}4 u_xD^2+\tfrac58(5u_{xx}+3u^2)D
+\tfrac{15}{16}(u_{xxx}+2uu_x).
$$

A solution $L$ of (7.11) is said to be {\it self-similar} if
$$
u_j(x,t)=\l^{n-j}u_j(\l x, \l^k t)\,,\qquad\l>0\,,\quad j=0,1,\dots n-2\,.
\tag7.12
$$
\noindent{\bf Examples.} The simplest cases are the self-similar KdV 
and Boussinesq solutions which satisfy, respectively:
$$
(xu)_x+u+\tfrac14 u_{xxx}+\tfrac32 u u_x=0\,;  
$$
$$
(xu_1)_x=u_{1xx}-2u_{0x}\,,\quad (xu_0)_x+u_0=\tfrac23u_{1xx}+
\tfrac23 u_1u_{1x}-u_{0xx}\,.
$$
The next equation in the KdV hierarchy leads to
$$
(xu)_x+u+\tfrac1{16}D^5u+\tfrac58 uu_{xxx}+\tfrac54u_xu_{xx}+\tfrac{15}8u^2u_x
=0.
$$

Note that (7.11) is the compatibility condition for the system of two
scalar equations
$$
Lv(x,t,z)=z^n v(x,t,z)\,,\qquad \pd{v}{t}=[L^{k/n}]_+ \,v\,.\tag7.13
$$
The self-similarity condition (7.12) is equivalent to a pair of operator
identities analogous to (7.6):
$$
[L-z^n I]\,\tl=\l^n\tl\,[L-z^n I]\,,\qquad
\L\,\tl=\l^k\tl\,\L\,,\tag7.14
$$
where
$$
\tl v(x,t,z)=v(\l x, \l^k\,t,\l\inv z)\,.\tag7.15
$$
The self-similarity condition is compatible with the scaling
condition
$$
v(\l x,\l^k t,\l\inv z)=v(x,t,z)\,,\qquad\l>0\,,\tag7.16
$$
on solutions of (7.15).  

We study the scalar equation $Lv=z^nv$ by using the standard procedure
to convert it to a first-order system.  Let
$$
\psi =\bigl(v,\pd{v}{x},\dots,\frac{\partial^{n-1} v}{\partial x^{n-1}}
\bigr)^t\,.
$$
Then $Lv=z^nv$ is equivalent to
$$
\pd{\psi}{x}=(J_z+q)\psi\,\tag7.17
$$
where
$$
J_z=\lmat 0&1&0&\dots&0\\0&0&1&\dots&0\\ & & \ddots &\\0&0&0&\dots&1\\
    z^n&0&0&\dots&0\rmat\,,\qquad
-q=\lmat 0&0&0&\dots&0\\0&0&0&\dots&0\\ & & \ddots &\\u_0&u_1&u_2&\dots&0\\
    \rmat\,.\tag7.18
$$
The commutator $[\,(L^{k/n})_+\,,\,L\,]$ is a differential operator of
order $n-2$:
$$
[\,(L^{k/n})_+\,,\,L\,]=\sum^{n-2}_{j=0}w_j\biggl(\frac{d}{dx}\biggr)^j\,.
\tag7.19
$$
If $Lv=z^nv$ then derivatives of $(L^{k/n})_+\,v$ with respect to $x$
can be expressed in terms of derivatives of $D^jv$ of order less than $n$,
i.e. the entries of $\psi$.  This leads to an identity
$$
(\,\L v,D\L v, \dots, D^{n-1}\L v\,)^t = G_k(x,z)\,\psi\tag7.20
$$
where the $n\times n$ matrix $G_k=G_{n,k}$ is a polynomial in $z^n$ and
also in the $u_j$ and their derivatives of order less than $k$; 
(7.20) holds for all such solutions $v$.  
\ss
\noindent{\bf Examples.} The two simplest cases are
$$
G_{2,3}=J_z^3 +\tfrac14\lmat -u_x&2u\\-2z^2u-2u^2-u_{xx}&u_x\rmat ;
$$
$$
G_{3,2} = J_z^2+\tfrac13\lmat 2u_1&0&0\\-3u_0+2u_{1x}&-u_1&0\\
2u_{1xx}-u_{0x}&-3u_0+u_{1x}&-u_1\rmat.
$$ 

We differentiate (7.20) with respect to $x$
to find that
$$
\biggl[\,\pd{}{x}-J_z-q\,,\,G_k\,\biggr]=r\tag7.21
$$
where $-r$ is the matrix corresponding to (7.19), i.e.
$$
-r = \lmat 0&0&0&\dots&0\\0&0&0&\dots&0\\ & & \ddots &\\w_0&w_1&w_2&\dots&0\\
    \rmat\,.\tag7.22
$$
It follows that the Gel'fand-Dikii equation (7.11) is the same as the 
matrix equation
$$
q_t=r.\tag7.23
$$

With this preparation we turn to the self-similar solutions of the
Gel'fand-Dikii equation.
The self-similarity condition (7.12) is equivalent to 
$$
q(\l x,\l^k t)=\l\inv d(\l)\inv q(x,t)d(\l)\,,\tag7.24
$$
where $d(\l)=\text{diag}(1,\l,\l^2,\dots,\l^{n-1})$.

If a scalar function  $v$ is invariant under the dilations (7.15),
the corresponding column vector $\psi$ transforms by
$$
\psi(x,t,z)=d(\l)\,\psi(\l x, \l^k t, \l\inv z)\,.\tag7.25
$$

Let $P$ denote the diagonal matrix
$$
P=\frac{d}{d\l}[d(\l)]_{\l=1}=\text{diag}(0,1,2,\dots,n-1)\,.\tag7.26
$$
The Euler equation equivalent to (7.24) is
$$
x\,\pd{q}{x}+kt\,\pd{q}{t}+[\,P\,,\,q\,]+q=0\,.\tag7.27
$$

The following result is analogous to Theorem 7.1 and can be proved in
the same way.

\proclaim{\bf Theorem 7.2}  If the matrix $q(x,t)$ corresponds to a 
self-similar
solution of the Gel'fand-Dikii equation (7.11), then $q(x,1/k)$ is a
solution of the system of algebraic ordinary differential equations 
$$
(xq)_x+[\,P\,,\,q\,]+r=0\,.\tag7.28
$$
Conversely, the self-similar extension of a  solution $q_0(x)$ of (7.28) 
corresponds to a solution of (7.11).\endproclaim

\noindent{\bf Examples.} The simplest cases are the self-similar KdV 
and Boussinesq solutions which satisfy, respectively:
$$
(xu)_x+u+\tfrac14 u_{xxx}+\tfrac32 u u_x=0\,;  
$$
$$
(xu_1)_x=u_{1xx}-2u_{0x}\,,\quad (xu_0)_x+u_0=\tfrac23u_{1xx}+
\tfrac23 u_1u_{1x}-u_{0xx}\,.
$$

\proclaim{\bf Proposition 7.3}  Equation (7.28) is equivalent to the 
commutator condition
$$
\biggl[\,\pd{}{x}-J_z-q\,,\,z\,\pd{}{z}-x(J_z+q)-P-G_k\,\biggr]=0\,\tag7.29
$$\endproclaim

\demo{Proof}  This is a straightforward calculation using (7.22) and
the identity 
$$
[\,P\,,\,J_z\,]-z\,\pd{}{z}(J_z)+J_z=0\,.\tag7.30
$$
\enddemo
We show in the next section that (7.29) is the equation for a 
monodromy-preserving flow.
\ms
{\bf 8. Gel'fand-Dikii equations and isomonodromy}
\ss 
The equation (7.29) which characterizes self-similar solutions of the
Gel'fand-Dikii equation (7.11), is the compatibility condition for the system
$$\align
\pd{\psi}{x}&=[J_z+q]\,\psi\,,\tag8.1\\
z\,\pd{\psi}{z}&=[xJ_z + xq+P+G_k]\,\psi\,.\tag8.2\endalign
$$

The basic result for the forward monodromy problem is the
following analogue of results of \S\S 2, 3.

\proclaim{\bf Theorem 8.1} The system (8.2) has a regular singular point
at $z=0$; the Stokes matrices for the irregular singular point at $z=\infty$
are preserved under the flow (7.28).
\endproclaim

The fact that the origin is a regular singular point for the system (8.2)
is a consequence of the fact that $G_k$ is a polynomial in $z$ (in fact a
polynomial in $z^n$).  To prove that the Stokes matrices are constant we
must analyze the behavior of solutions of (8.2) as $z$ tends to infinity.
The argument is similar to that in \S 2, and needs some preparation.

Note first that in the trivial case $q=0$ there is a fundamental
solution of (8.1) having the form $\psi=\Lambda_ze^{xzJ}=d(z)\Lambda 
e^{xzJ}$ where
$$
\Lambda=\lmat 1 & 1 & \dots & 1\\ \a_1 & \a_2 & \dots &\a_n\\
                &   & \ddots & \\
              \a_1^{n-1}&\a_2^{n-1}&\dots &\a_n^{n-1}\rmat\,,\qquad
J=\text{diag}(\a_1,\a_2\,\dots,\a_n)\,,\tag8.3
$$
and the $\a_j$ are the $n$--th roots of unity; to fix the choice we
let $\a_j=\exp(2\pi ij/n)$. Moreover
$$\gather
\Lambda_z\inv\,J_z\,\Lambda_z = zJ\,;\tag8.4\\
\tilde q\equiv\Lambda_z\inv\,q\,\Lambda_z = O(z\inv)\,,\qquad
\tilde r\equiv\Lambda_z\inv\,r\,\Lambda_z = O(z\inv)\,.\tag8.5
\endgather$$
It is natural to look for solutions to (8.2) in the form $\psi=\Lz me^{xzJ}$,
so (8.1) and (8.2) become
$$\align
\pd{m}{x}&=[\,zJ\,,\,m\,]+\tilde q m\,.\tag8.6\\
z\,\pd{m}{z}&=[xzJ+x\tilde q+\tilde G]m 
-m[(zJ)^k+xzJ],\qquad \tilde G=\Lz\inv G\Lz\tag8.7
\endalign
$$

Equation (8.6) has formal solutions as in \S 1: $m=\sum^\infty_{j=0}
z^{-j}f_j$ with $f_0=\pmb 1$. 

\noindent{\bf Example.} With $n=2$ and with $x_0$ fixed, the unique formal 
solution with diagonal part $\equiv \pmb 1$ at $x=x_0$ is 
$$\align
m(x_0,z)=\pmb 1 &+(\tfrac14 z^{-2} u+\tfrac1{16}z^{-4}
[u_{xx}+2u^2])\lmat 0&1\\1&0
\rmat\\
&+(\tfrac18 z^{-3}u_x+\tfrac1{32}z^{-5}[u_{xxx}+6uu_x])\lmat 0&-1\\1&0\rmat
+O(z^{-6}).
\endalign
$$

For later use we note a symmetry property of (8.6).  Let $\Pi$ be
the permutation matrix
$$
\Pi = J_1=\lmat 0&1&0&\dots&0\\0&0&1&\dots&0\\ \ddots \\1&0&0&\dots&0
\rmat.
$$
Then simple calculations show 
$$
\Pi\,(zJ)\,\Pi\inv=(\a z)J,\quad \Pi\, \tilde q_z\,\Pi\inv=\tilde q_{\a z},\quad
\Pi\, \tilde r_z\,\Pi\inv=\tilde r_{\a z},
\quad \Pi\,\Lz\,\Pi\inv=\Lambda_{\a z}.\tag8.8
$$

Note that $\tilde G_k=\Lambda_z\inv\, G_k\,\Lambda_z$ is a rational
function of $z$ with leading term $z^kJ^k$ as $z$ tends to infinity.

\noindent{\bf Example.} With $n=2$ and $k=3$, 
$$
\tilde G_{3}=z^3 J +\tfrac12 zu\lmat 0&1\\-1&0\rmat -\tfrac14 u_x
\lmat 0&1\\1&0\rmat+\tfrac 18 z\inv (u_{xx}+ 2u^2)\lmat 1&1\\-1&-1\rmat.
$$

The next step in our analysis of the behavior of solutions for large $z$
involves understanding $\tilde G_k$.

\proclaim{\bf Lemma 8.2} If $\hat m$ is a formal solution of (8.6),
then the product $\hat m\,(zJ)^k\,\hat m\inv$  differs from $\tilde G_k$
only in terms of negative degree in $z$.  \endproclaim

\demo{Proof} Equations (7.21) and (8.6) 
imply that
$$
\biggl[\,\pd{}{x}-zJ\,,\,\hat m\inv\tilde G_k \hat m\,\biggr]=
\hat m\inv \tilde r \hat m=O(z\inv)\,.
$$
We write
$$
\hat m\inv\tilde G_k \hat m=\sum^\infty_{j=0}z^{k-j}g_j,\qquad 
\hat m\inv \tilde r \hat m =\sum^\infty_{j=0}z^{-j}h_j,\tag8.9
$$
and take $g_j=h_j=0$ for $j<0$, so that the equation above is equivalent to 
$$
[\,J\,,\,g_{j+1}\,]=(g_j)_x-h_{j-k},\qquad\text{all}\ \ j.\tag8.10
$$
It follows from the relations (8.10) and the identity $g_0=J^k$ that
$g_j$ is diagonal and constant for $j\leq k$.  Note that this fact does not
depend on (8.2); it is true for {\it arbitrary} smooth $q$.
As in \S 1, the fact that the diagonal part of $\hat m\inv\tilde G_k \hat m$ is
independent of the choice of formal solution $\hat m$ implies that the
coefficients of the diagonal part are polynomials in the entries
of $q$ and their derivatives, and that all but the top order term
$(zJ)^k$ are polynomials with constant term zero.  The only such polynomial
which is constant for arbitrary $q$ is the zero polynomial.
Therefore 
$g_j=0$ for $0<j\leq k$ and the assertion
is proved.\enddemo

As in \S 2 we may use a partial sum of a formal solution to regularize our
problem at infinity.
Let $f=\sum^k_{j=0} z^{-j}f_j$ be a partial sum of $\hat m$; we have
$$
f\inv\,\tilde G_k\,f = z^kJ^k+O(z\inv)\,.
$$
We look for a solution of (8.2) in the form $\psi=\Lambda_z\,f\,\hat\psi$.
Note that
$$
z\,\pd{\psi}{z}-P\,\psi=\Lambda_z\,z\,\pd{}{z}\bigl(\Lz\inv \psi\bigr)\,
\tag8.11
$$
so the equation for $\hat\psi$ is
$$\align
z\,\pd{\hat\psi}{z}&=[-z f\inv f_z+xf\inv (zJ+\tilde q) f+
f\inv\tilde G_k f]\,\hat\psi \tag8.12\\
&=[z^kJ^k+xzJ+r(x,z)]\,\hat\psi\,,\quad\qquad r(x,z)=O(z\inv)\,.\endalign
$$
From this point the analysis is essentially the same as in \S 2.
Equation (8.2) has unique solutions of the form
$$
\pn=\Lz\mn e^\Phi \,,\qquad \Phi(x,z)=\tfrac1k z^kJ^k+xzJ\,,
\tag8.13
$$
where $\mn$ has an asymptotic expansion with
leading coefficient $\pmb 1$, valid in a sector $\On$.  As in \S 2 these
functions also satisfy (8.1) and therefore the Stokes matrices which
relate the $\pn$ are constant.  This completes our sketch of the proof
of Theorem 8.1.
\ss
\noindent{\bf Remark.} The symmetry property (8.8) and the uniqueness of the 
solutions (8.13) allow us to conclude that these solutions have the
n-fold symmetry which can be stated roughly as
$$
m(x,\a z)=\Pi\, m(x,z)\,\Pi\inv.
$$
To state it more precisely we need to investigate the set
$$
\Sigma =\{z: \text{Re}(z\a_j)^k=\text{Re}(z\a_l)^k, \text{ some } \a^k_j\neq \a^k_l\}.
$$
Let $p$ be the smallest positive integer such that $\a^{pk}=1$, i.e. such
that $n$ divides $pk$.  Then
$$
\Sigma=\{z: z^{k}\in i\Bbb R\}\ \text{ if } p=2;\qquad \Sigma=\{z: z^{2pk}\in
\Bbb R_+\}\ \text{ if } p>2.
$$
We define regions $\On$ as in \S 2, bounded by various of the rays of the
set $\Sigma$.  It follows from the form of $\Sigma$ that 
$\Omega_{\nu+1}$ is obtained by rotation of $\On$ through 
an angle $\pi\,/pk$, ($\pi/k$ if $p=2$) and the precise form of the symmetry is
$$
m_{\nu+s}(x,\a z)=\Pi\,\mn(x,z)\,\Pi\inv,\qquad \cases s=2kp/n & \text{if}\ 
p>2 \\ s=2k/n & \text{if}\ p=2. \endcases
\tag8.14
$$
The Stokes matrices satisfy the corresponding symmetry
$$
S_{\nu+s}=\Pi\, \Sn\,\Pi\inv.\tag8.15
$$
as well as the standard constraints
$$
\text{diag}(\Sn)=1,\qquad e^\Phi \Sn e^{-\Phi} \text{ is bounded as } z\to
\infty,\ \ z\in\On,\ \  z \text{ near } \Sigma_{\nu},\tag8.16
$$
where $\Sigma_{\nu}$ is any ray in $\On\cap\Omega_{\nu+1}$; it will be
convenient to choose $\Sigma_{\nu}$ to bisect this sector. 
\ss
The forward monodromy problem at $z=0$ is the same as in \S 3, but
we must take the symmetry into account.
For convenience we consider only the generic case, when there as a 
traceless constant matrix $A$ and a solution of $\psi$ of (8.1),
(8.2) with the 
property that $\psi(x,z)z^{-A}$ is an entire function of z, invertible for
each $z$.  
If so, then there are traceless constant matrices $\Bn$ such that
$\pn z^{-\Bn}$ are entire, and
$$
B_{\nu+1}=\Sn\inv \Bn\Sn;\qquad \exp(2\pi\, i\Bn)=\Sn S_{\nu+1}\dots S_{\nu-1},
\qquad B_{\nu+s}=\Pi\, \Bn\,\Pi\inv.\tag8.17
$$

\proclaim{\bf Theorem 8.3} In the generic case, if two solutions of the
isomonodromy equation (7.29) have a common domain and the same monodromy
data $\{\Sn,\Bn\}$, then they are identical.
\endproclaim

\demo{Proof} Suppose that $q$ and $q'$ are two solutions of (7.29) on
a common domain, having the same monodromy data.  Let $\pn$ and 
$\pn'$ be the corresponding normalized solutions of (8.1), (8.2). Let
$m=\pn e^{-\Phi}$ for $|z|>1$, $z$ between $\Sigma_{\nu-1}$ and $\Sigma
_{\nu}$  and $m=\pn z^{-\Bn}e^{-\Phi}$
on $\{|z|<1\}$.  Define $m'$ analogously.  Each of $m$ and $m'$ is a
solution of a matrix Riemann-Hilbert factorization problem; a vector
version of this problem 
is described in more detail below.  It follows that $m\inv m'$
is continuous and piecewise holomorphic, hence entire.  From the
asymptotics of (8.13) we know that $\Lz\inv m\sim \pmb 1$ at infinity,
and the same for $m'$, so  $Q=m\inv m'$ is a polynomial in $z$ and
$$
d(z)\inv Q d(z)=\pmb1+O(z\inv).
$$
This last fact implies that $Q-\pmb1$ is strictly lower triangular.
Therefore $m'=Qm$ has the same first row as $m$.  But $m$ and $m'$
are each determined by the first row $M$; in fact the $j$-th row is
$D^{j+1} (Me^\Phi)\,e^{-\Phi}$.  Thus $\psi=\psi'$ and $q=[D\psi_-J_z\psi]
\psi\inv=q'$.\qed 
\enddemo   

In order to formulate the inverse problem, in which matrices
$\{\Sn,\Bn\}$ are given, we must give a precise description of the vector
Riemann-Hilbert problem alluded to above.   
Let $M$ be the first row of $\pn e^{-\Phi}$
for $|z|>1$, $z$ between $\Sigma_{\nu-1}$ and $\Sigma_{\nu}$.   
On $\{|z|<1\}$ let $M$ be the first row of
$\pn z^{-\Bn}e^{-\Phi}=\Lz\mn$; this is independent of $\nu$.  Because
of the asymptotics of $\mn$ and the relations given by the Stokes matrices,
one can see that the row vector function $M$ has the properties
$$\align
&M \text{ has limit } (1,1,\dots,1) \text{ as } z\to\infty;\tag8.18\\
&M \text{ is bounded and holomorphic where defined;}\tag8.19\\
&M \text{ is continuous up to the boundary from each component;}\tag8.20
\endalign
$$
Moreover the boundary values $M_\nu$, the limit on the boundary of the
region outside the unit circle between the rays $\Sigma_{\nu-1}$ and
$\Sigma_{\nu}$, 
and $M_\Gamma$, the limit on the unit circle from the disc, satisfy
$$
M_{\nu+1}=M_{\nu}e^{\Phi}\Sn e^{-\Phi};\qquad M_{\nu}=M_{\Gamma}e^\Phi z^{\Bn}
e^{-\Phi},\tag8.21
$$
while the row vector $M$ itself has the symmetry
$$
M(x,\a z)=M(x,z)\,\Pi\inv,\qquad |z|>1,\ \ \ z\notin\bigcup \Sigma_{\nu}.
\tag8.22
$$

Conversely, suppose matrices $\{\Sn,\Bn\}$ are given.
Generically the Riemann-Hilbert problem (8.18)-(8.22) has a unique
solution, say for $x$ in some domain.  

\proclaim{\bf Theorem 8.4} Suppose that $\Sn$ and $\Bn$ are constant matrices 
which satisfy the conditions (8.15)-(8.17), and $\tr \Bn=0$.  
Suppose the Riemann-Hilbert
problem (8.18)-(8.22) has a unique solution $M(x,\cdot)$, for $x$ in 
some domain.  Then there is a unique $n$-th order operator
$$
L=D^n+\sum^{n-2}_{j=0}u_j(x)D^j,\qquad D=\frac{d}{dx}
$$
such that the row vector $v=Me^\Phi$ satisfies the
equation $Lv=z^nv$.  The system corresponding to $L$ satisfies the
isomonodromy equation (7.29).\endproclaim

\demo{Proof}
The vector function $M$ has an asymptotic expansion in powers of $z\inv$
as $z\to\infty$,
$$
M(x,z)\sim \sum^\infty_{j=0}z^{-j}a_j(x)\tag8.23
$$
and the symmetry (8.22) implies that the coefficients have the form
$$
a_j(x)=b_j(x)(1,1,\dots,1)J^{-j},\tag8.24
$$
where the $b_j$'s are scalars. The expansion can be differentiated term
by term with respect to $x$.    

Next we construct the sequence of vectors
$$
M^{(j)}=D^j(Me^\Phi)e^{-\Phi}=(D^jv)e^{-\Phi},\quad j\geq 0.\tag8.25
$$
Note that 
$$
M^{(j)}=(1,1,\dots,1)(zJ)^j + O(z^{j-1}).\tag8.26
$$
In particular $M^{(n)}-z^nM$ is $O(z^{n-1})$.  
We use the symmetries (8.24) and their analogues for 
the $M^{(j)}$'s to conclude that there is a unique scalar function
$u_{n-1}(x)$ such that
$$
M^{(n)}-z^nM+u_{n-1}M^{(n-1)}=O(z^{n-2}).\tag8.27
$$
Continuing, we find unique functions $u_{n-2},\dots u_0$ such that
$$
M^{(n)}-z^nM+\sum^{n-1}_{j=0}u_j M^{(j)}=0(z\inv).\tag8.28
$$
Now the $M^{(j)}$'s  and $z^nM$ 
are solutions of the Riemann-Hilbert problem
(8.19)-(8.22).  Therefore the left side of (8.28) is a solution,
with limit $0$ as $z$ tends to $\infty$.
By our uniqueness assumption for solutions of (8.18)-(8.22) the left
side of (8.28) must vanish.  It follows that the row vector $v=Me^{\Phi}$ 
satisfies the equation
$$
Lv\equiv [D^n+\sum^{n-1}_{j=0} u_jD^j]v=z^nv.\tag8.29
$$
Our next task is to show that $u_{n-1}\equiv 0$, so that $L$
has the desired form.

We may reduce to a first-order system as before and conclude that
$v$ is the first row of a solution to a system of the form
(8.1).  The $j$-th row of the matrix $\psi$ is precisely $M^{(j-1)}e^\Phi$.
Note that the trace of the matrix $q$ is $-u_{n-1}$.  It follows that the
determinant $\bigtriangleup=\det(\psi)=\det(\psi e^{-\Phi})$ satisfies 
$$
\bigtriangleup_x\equiv-u_{n-1}\bigtriangleup.\tag8.30
$$
On the other hand, $\bigtriangleup$ is a solution of a scalar 
Riemann-Hilbert problem with trivial multiplicative jumps.  Therefore 
$\bigtriangleup$ is an
entire function.  The asymptotic behavior is the same as that of 
$\det\Lz$, so $\bigtriangleup$ is a polynomial with leading coefficient
$\det\Lz$.  Since this coefficient 
does not depend on $x$, (8.30) implies $u_{n-1}\equiv
0$.

Similarly, $zM_z-xM^{(1)}+M[(zJ)^k+xzJ]-M^{(k)}$ is a solution of 
(8.19)-(8.22) 
which is $O(z^{k-1})$ and we may use the symmetries and (8.26) to deduce that
$$
zM_z-xM^{(1)}+M[(zJ)^k+xzJ]=M^{(k)}+a_{k-1}M^{(k-1)} +O(z^{k-1}).
$$
As before, we conclude that $v$ satisfies an equation of the form
$zv_z=xv_x+L^\#v$, where $L^\#$ is a differential operator in the $x$-variable,
of order $k$.  This equation is equivalent to the first-order system
$$
z\psi_z=x(J_z+q)\psi + P\psi+ G\psi,
$$
where the matrix $G(x,z)$ is characterized by the condition 
analogous to (7.20):
$$
(\,L^\# v,DL^\# v, \dots, D^{n-1}L^\# v\,)^t = G\psi.\tag8.31
$$
Equation (7.29) is the  compatibility
condition for (8.31) and (8.31), provided we may replace $L^\#$ by $\L$. 
 We may make this replacement if
and only if $L^\#v=\L v$, and this is true if the matrices
$G$ and $G_k$ have the same first row.

We write $\psi=\Lz m e^\Phi$ and note that $m$ has an 
expansion at $\infty$ with leading term $\pmb 1$.  Moreover,
$m$ satisfies
$$
xm_x=[\,xzJ\,,\,m\,]+x\tilde qm;\quad zm_z=[\,xzJ\,,\,m\,]+
[x\tilde q+\tilde G]m-m(zJ)^k,
$$
where $\tilde G=\Lz\inv G\Lz$.  It follows from this and Lemma 8.2 that
$$
\tilde G=m(zJ)^km\inv+xm_x-zm_z=m(zJ)^km\inv+O(z\inv)=\tilde G_k+O(z\inv).
\tag8.32
$$
Therefore 
$$
d(z)\inv[G-G_k]d(z)=\Lambda[\tilde G -\tilde G_k]\Lambda\inv=O(z\inv)
$$
which implies that $G-G_k$ is strictly lower triangular.  In particular $G$
and $G_k$ have the same first row.  This in turn
implies that $L^\#v=\L v$, so we may replace $L^\#$ by $\L$. \qed
\enddemo

\ms
{\bf 9.  Isomonodromy deformations and string equations}
\ss
In this section we consider a second classes of (systems) of ordinary 
differential
equations associated to the Gel'fand-Dikii hierarchy (7.11) and show that
each is an isomonodromy deformation for a first order system.  This class
was obtained by M. Douglas
[Do] in connection with theories of quantum gravity.  The
Gel'fand-Dikii flow itself is replaced by the equation
$$
[(L^{k/n})_+\,,\,L\,]=\hbar I\tag9.1
$$
where $I$ is the identity operator.  For convenience of notation we 
consider $\hbar=1$.  The case $n=2$, $k=3$ and
the case $n=3$, $k=2$ each lead to a form of the PI equation; for
$n=2$, $k=3$ the equation is
$$ 
\tfrac14 u_{xxx}+\tfrac32 u u_x +1=0\,;
$$
for $n=3$, $k=2$ the constants are different.  The case $n=2, k=5$
is 
$$
\tfrac1{16}D^5u+\tfrac58 uu_{xxx}+\tfrac54 u_xu_{xx}+\tfrac{15}8 u^2u_x
+1=0.
$$
\ss
\noindent{\bf Remark.} The development in this section applies equally
to the case when the operator $\L$ is replaced by a finite linear
combination $\sum_k a_k\L$ with arbitrary constants $a_k$.  The phase
function $\Phi$ which is considered below must be changed accordingly.
\ss
Note  that (9.1) can be written as a comutator equation
$$
\bigl[\,L-z^nI\,,\,\pd{}{z}-nz^{n-1}(L^{k/n})_+\,\bigr]=0\,,\tag9.2
$$
which is the compatibility condition for the system of two scalar
equations
$$
Lv=z^nv,\qquad nv_z=nz^n\L v.\tag9.3
$$

Reduction to a first order system as in \S 8 gives a matrix version
of (9.2):
$$
\biggl[\,\pd{}{x}-(J_z+q)\,,\,z\,\pd{}{z}-nz^n G_k\,\biggr]=0\,.\tag9.4
$$
In turn, (9.4) is the compatibility condition for the system 
$$\align
\pd{\psi}{x}&=[J_z+q]\,\psi\,,\tag9.5\\
z\,\pd{\psi}{z}&=nz^n G_k\,\psi\,.\tag9.6
\endalign
$$

We sketch a proof of the following analogue of Theorem 8.1.

\proclaim{\bf Theorem 9.1} The system (9.6) has a regular point
at $z=0$; the Stokes matrices for the irregular singular point at $z=\infty$
are preserved under the flow (9.4).\endproclaim

Since $G_k$ is a polynomial in $z^n$
it follows that the origin is a regular point for (9.6).
The second assertion is a consequence of
Lemma 9.3 below.  We look for exact solutions of (9.6) having the
form $\psi=\Lz f\hat\psi$, where $f=\sum^{n+k+1}_{j=0}z^{-j}m_j$.  Then
as before there are solutions $\pn$ of this form which have the appropriate
asymptotic expansions in sectors $\On$, and which are related by 
constant Stokes matrices $\Sn$.  

Thus we look for solutions of (9.5), (9.6) having
the form $\psi = \Lz me^\Phi$ for some diagonal matrix-valued function $\Phi$.
The equations become
$$\align
D_x m&\equiv m_x-zJm-q_zm=-m\Phi_x\;\tag9.7\\
D_zm&\equiv zm_z-nz^n\tilde G_k m +\Lz\inv P\Lz m = -zm\Phi_z\,.\tag9.8
\endalign
$$

\proclaim{\bf Lemma 9.2} Suppose $\hat m$ is a formal solution of (9.7).
Then the diagonal part of the coefficient $g_j$ of $z^{-j}$ in the 
formal series
$\hat m\inv r_z \hat m$ has the form $c_j J^{-j}$, where $c_j$ is a 
scalar function which is a polynomial without constant term in the 
entries of $q$ and their derivatives.  Moreover, $c_j=0$ if $j$ is divisible
by $n$.  \endproclaim

The proof is postponed.

\proclaim{\bf Lemma 9.3} Suppose (9.1) is satisfied.  Then there
is a unique diagonal matrix-valued function
$$
\Phi(x,z)=\tfrac{n}{n+k}(zJ)^{n+k}+\sum_{j=1}^{n-1}b_j(zJ)^j+
xzJ-\tfrac12 (n-1)\log z\,\pmb 1\,,\tag9.9
$$
where the $b_j$'s are scalar constants, and a unique formal
power series in $z\inv$, $m=\sum^\infty_{j=0}z^{-j}m_j(x)$ such that
$m_0=\pmb 1$ and
$m$ is a formal solution of the equations (9.7), (9.8).\endproclaim

\demo{Proof} We begin by refining Lemma 8.2.
Let $\hat m$ be a formal solution of (9.5) and use the notation of (8.9).

We already know from the proof of Lemma 8.2 that $g_1=\dots=g_k=0$.
It follows from Lemma 9.2 and the recursion relations (8.10) that
the diagonal part of $g_{k+j}$ has derivative $c_jJ^{-j}$ for $j>0$.
The diagonal part of $g_{k+j}$ itself is a polynomial without constant
term in the entries of $q$ and their derivatives, so we can write
$$
(g_{k+j})^{\text{diag}}=\tfrac{n-j}{n}b_{n-j}J^{-j}=\tfrac{n-j}{n}
b_{n-j}J^{n-j},\qquad j>0,
$$
where $b_j$ is scalar polynomial in the entries of
$q$ and their derivatives, without constant term.  

Equation (9.1), which we now invoke,
implies that $w_0=-1$ and $w_j=0$, $j>0$.  Therefore $h_j=0$ for $j<n-1$
and 
$$
(h_{n-1})^{\text{diag}}=-\tfrac1n w_0 J=\tfrac1n J.
$$
We use the recursion relations (8.10) again and
conclude that $g_{k+j}$ is diagonal and is constant for $0<j<n-1$.
Thus the scalar $b_j$ is constant, i.e. this expression in $q$ and
its derivatives is invariant under the $x$-flow (9.1).  

At the next step (8.10) implies
$$\align
g_{k+n-1}&=\tfrac1n (xJ + b_1J),\qquad b_1\quad\text{constant, scalar};\\
(g_{k+n})_x&=[\,J\,,\,g_{k+n+1}\,]+h_n.
\endalign 
$$ 
By Lemma 9.2, $h_n$ is off-diagonal and we find 
that the diagonal part of $g_{k+n}$ is constant. This fact 
is independent of (9.1) and holds for arbitrary $q$, which again allows
us to conclude that the diagonal part of $g_{k+n}$ vanishes in all cases.

Note that the diagonal part of $\Lambda_z\inv P\Lambda_z=\Lambda\inv P\Lambda$
is $\tfrac12 (n-1)\,\pmb 1$.  
The preceding argument shows that 
$$\align
\hat m\inv [nz^n&\tilde G_k -\Lambda_z\inv P\Lambda_z]\hat m\tag9.10\\
&=nz^{n+k}J^k+\sum^{n-1}_{j=1}j b_{j}(zJ)^j +xzJ -\tfrac12 (n-1)\,\pmb 1+E\\
&=z\pd{}{z}\bigl[\tfrac{n}{n+k}z^{n+k}+\sum^{n-1}_{j=1}b_j(zJ)^j+xzJ-
\tfrac12 (n-1)\log z\,\pmb 1\bigr]+E\\
&=z\pd{}{z}\bigl[\Phi (x,z)\bigr]+E\endalign
$$
where the $b_j$ are scalar constants and  $\Phi$ is defined by (9.9).
The remainder term $E$
is a formal series with terms of non-positive degree in $z$, whose
term of degree $0$ in $z$ is off-diagonal.  

Finally we look for a solution to (9.8) in the form $m=ph$ where $p$ 
is the sum 
of the terms of degree $\geq -n-k$ in the formal solution $\hat m$.
An analogue of (9.10) holds with $p$ in place of $m$, so (9.8) is
equivalent to
$$
zh_z=[\,\Phi_z\,,\, h\,] -Fh\,,\qquad F=O(z\inv)\,
$$
where the off-diagonal part of $F$ is $O(z^{-2})$.  The proof 
of Theorem 2.1 in [CL, Ch. 5] shows that this equation has a formal solution
$h=\pmb 1 +O(z\inv)$.  
\enddemo

We return to unfinished business.

\demo{Proof of Lemma 9.2}  The assertion is independent of the choice of
formal solution, so we may take $\hat m$ to be the unique formal solution
normalized so $\hat m(x_0,z)\equiv\pmb 1$.  Because of uniqueness and the
symmetry properties (8.8), we can conclude that $\Pi\,\hat m(x,z)\,\Pi\inv\equiv
\hat m(x,\a z)$.  With this and another use of (8.8) we find that
$$
\Pi\,\hat m(x,z)\inv r_z\hat m(x,z)\,\Pi\inv
=\hat m(x,\a z)\inv r_{\a z}\hat m(x,\a z).
$$
Therefore the coefficients in the expansion $\hat m\inv r_z\hat m=\sum
z^{-j}h_j$ satisfy $\Pi\, h_j\,\Pi\inv=\a^{-j} h_j$.  This in turn implies
that 
$$
\Pi\,(J^j h_j)\,\Pi\inv=(\Pi\, J^j \,\Pi\inv)(\Pi\, h_j\,\Pi\inv)=(\a^j J^j)(\a^{-j}h_j)
=J^j h_j.
$$
It follows that the diagonal part of $J^j h_j$ commutes with $\Pi\,$ and 
therefore
is a scalar $c_j$.  We noted earlier that since the diagonal part of $h_j$
is independent of the choice of $\hat m$, its entries are polynomials
in the entries of $q$ and their derivatives.  

When $j$ is divisible by $n$, the diagonal part of $h_j$ has trace $nc_j$,
so to complete the proof it is enough to show that the trace is zero.  But 
$$
\tr(\hat m\inv r_z\hat m)=\tr(r_z)=\tr(\Lambda_z\inv r\Lambda_z)=\tr(r)\equiv
0.
$$\enddemo

\noindent{\bf Examples.}  1. In view of the previous calculations for the case 
$n=2$, $k=3$, we find that in general
$$
\hat m\inv (2z\tilde G_{3})\hat m =2z^4J^3+\tfrac18 \lmat 2u_{xx}+6u^2
 & z\inv (u_{xxx}+6uu_x)\\ \ z\inv(u_{xxx}+6uu_x)
& -2u_{xx}-6u^2\rmat.
$$
Equation (9.1) in this case is $u_{xxx}+6uu_x=-4$, so $u_{xx}+3u^2=-4(x+b)$
for some constant $b$.  Thus the general formula reduces to
$$
\hat m\inv(2z\tilde G_3)\hat m=2z^4J^5+(x+b)J-\tfrac12 z\inv
\lmat 0&1\\1&0\rmat.
$$ 
It follows that for this case
$$
\Phi(x,z)=\tfrac25 (zJ)^5 + (x+b)zJ - \tfrac12 \log z\,\pmb 1.
$$

2. For the case $n=3$, $k=2$, in general 
$$
[\hat m\inv(3z^2\tilde G_2)\hat m]^{\text{diag}}=3z^4J^5 +(u_{1xx}-2u_0)zJ^2+
(\tfrac23 u_{1xx}+\tfrac13 u^2_1-u_{0x})J-z\inv\pmb 1.
$$
Equation (9.1) is equivalent to: $u_{1xx}-2u_0\equiv 2b_2$ and
$\tfrac23 u_{1xx}+\tfrac13 u^2_1-u_{0x}=b_1+x$ for some constants $b_2$,
$b_1$.  Therefore
$$
\Phi(x,z)=\tfrac35 (zJ)^5 + b_2(zJ)^2 + (b_1+x)zJ-\log z\,\pmb1.
$$ 

\noindent{\bf Remark.} The set of solutions of (9.1) is invariant under 
translation, so one can always get rid of the constant $b_1$ in $\Phi$ 
by translating.
\ss
\noindent{\bf Remark.} As we noted earlier, the origin is a regular point 
for the 
equation (9.6), which is satisfied by $\psi=\Lambda_z me^\Phi$. 
Therefore $me^\Phi$ has trivial monodromy.
It follows that the product of the Stokes matrices is $(-1)^{n-1} \pmb 1$:
$$
S_1 S_2 \cdots S_0 = (-1)^{n-1}\pmb 1.\tag9.11
$$
\ss
The following is the analogue of Theorem 8.3, and is proved in the same
way.

\proclaim{\bf Theorem 9.4} If two solutions of the isomonodromy equation
(9.4) have a common domain and the same monodromy data $\{\Sn\}$, then
they are identical.\endproclaim

The direct problem is similar to, but somewhat simpler than, the
direct problem treated in \S 8.  The set $\Sigma$ has the form
$$
\Sigma =\{z: z^{n+k}\in i\Bbb R\}\ \text{ if } p=2;\qquad
\Sigma =\{z: z^{2pn+2pk}\in\Bbb R_+\}\ \text{ if } p>2,\tag9.12
$$
where again $p$ is the smallest integer such that $n$ divides $pk$.
The symmetry condition (8.8) and its consequences carry over to the
present situation, with $s=2p(n+k)/n$ if $p>2$ or
$s=2(n+k)/n$ if $p=2$..  

Again we associate a Riemann-Hilbert problem.  Let
$\pn=\Lz \mn e^\Phi$ with $\mn=\pmb 1+0(z\inv)$ in $\On$ and let
$M$ be the first row of $\mn$ for $|z|>1$, $z$ between $\Sigma_{\nu-1}$
and $\Sigma_{\nu}$; on the unit disc let $M$ be the first row of 
$z^{(n-1)/2}\psi_0e^{-\Phi}$.  Then $M$ satisfies (8.18)-(8.20), and (8.22), 
and its boundary values satisfy
$$\align
&M_{\nu+1}=\Mn e^\Phi \Sn e^{-\Phi},\qquad |z|>1;\tag9.13\\
&M_{\nu+1}=z^{-(n-1)/2}M_\Gamma\,e^{\Phi}S_0S_1\cdots S_{\nu}e^{-\Phi},
\qquad |z|=1.\tag9.14
\endalign
$$
 
For the inverse problem, suppose that the $\Sn$ and $\Phi$ are given.
Generically the Riemann-Hilbert problem
(8.18)-(8.20), (9.13), (9.14)
has a unique solution $M(x,\cdot)$ for $x$ in some domain.  

\proclaim{\bf Theorem 9.5} Suppose that the constant matrices $\Sn$ satisfy
(8.15), (8.16), and (9.11) and that the matrix $\Phi(x,z)$ has the form
(9.9).  Suppose that the Riemann-Hilbert problem
(8.18)-(8.20), (9.13), (9.14) has a unique solution for $x$ in some domain.
Then there is a unique $n$-th order operator $L=D^n+\sum^{n-2}_{j=0}u_jD^j$
such that $v=Me^\Phi$ satisfies the equations $Lv=z^nv$, $zv_z=nz^n\L v$.
In particular, the isomonodromy equation (9.1) is satisfied.
\endproclaim

\demo{Proof} Arguing exactly as in the proof of Theorem 8.4, we examine
the expressions
$$ 
M^{(n)}-z^nM,\qquad zM_z+M(z\Phi_z)-nz^nM^{(k)}
$$
and show that $v=Me^\Phi$
satisfies an equation of the correct form $Lv=z^nv$ as well as a second
equation of the form $zv_z=nz^nL^\# v$.  Here $L^\#$ is an operator of
order $k$.  The corresponding first order systems are
$$
\psi_x=[J_z+q]\psi,\qquad z\psi_z = nz^nG\psi,
$$
with $G$ determined by (8.31).  As before we write $\psi=\Lz me^\Phi$,
so $m=\pmb 1 +O(z\inv)$ and
$$
zm_z=[nz^n\tilde G-\Lz\inv P\Lz]\, m-m(z\Phi_z)
=[nz^n \tilde G-\Lambda\inv P\Lambda]\, m-m(z\Phi_z).
$$  
Because of the form of $\Phi$ and the asymptotic behavior of $m$, 
this equation implies
$$
\tilde G = m(zJ)^km\inv + O(z\inv)=\tilde G_k +O(z\inv).
$$
As in \S 8 we conclude from this fact that $G$ and $G_k$ have the 
same first row and therefore that we may replace $L^\#$ by $\L$.  
\ms

\noindent{\bf References}
\ss
[AS1]\ \ \ \  M. J. Ablowitz and H. Segur, 
{\sl Exact linearization of a Painlev\'e transcendent,}
Phys. Rev. Lett. {\bf 38} (1977), 1103-1106.
\ss 
[AS2]\ \ \ \  M. J. Ablowitz and H. Segur, 
``Solitons and the Inverse Scattering Transform'',
 SIAM Studies in Applied Mathematics,  Philadelphia 1981. 
\ss
[Ai]\ \ \ \  H. Airault,
{\sl Rational Solutions of Painlev\'e Equations,}
 Studies in Applied Mathematics {\bf 61} (1979), 31-53.
\ss
[BC]\ \ \ \ R. Beals and R. R. Coifman,
{\sl Scattering and inverse scattering for first order systems,}
Comm. Pure Appl. Math. {\bf 87} (l984), 39-90.
\ss
[CL]\ \ \ \ E. A. Coddington and N. Levinson,
``Theory of Ordinary Differential Equations'', McGraw-Hill, New York 1955.
\ss
[Do]\ \ \ \  M. Douglas, 
{\sl Strings in less than one dimension and the generalized 
KdV hierarchies,} Phys. Lett. {\bf 238B} (1990), 176-180.
\ss
[FN]\ \ \ \  H. Flaschka and A.C. Newell,
{\sl Monodromy and Spectrum Preserving Deformations. I,}
 Comm. in Math. Phys. {\bf 76} (1980), 65-116.
\ss
[FZ]\ \ \ \  A. S. Fokas and X. Zhou,
{\sl Integrability of Painlev\'e transcendents,} to appear.

\ss
[Ga]\ \ \ \  E. Gambier,
{\sl Sur les \'equations differentielles du second
ordre et du premier degr\'e dont
l'int\'egrale g\'en\'erale est \`a points critiques fixes,}
 Acta Math. {\bf 33} (1910), 1-55.
\ss
[IN]\ \ \ \  A. R. Its and V. Y. Novokshenov,
`` The Isomonodromic Deformation Method in the Theory
of Painlev\'e Equations,''
Lecture Notes in Mathematics no. 1191,
Springer Verlag, Heidelberg  1986.
\ss
[JMU]\ \  M. Jimbo, T. Miwa, and K. Ueno,
{\sl Monodromy preserving deformation of linear
ordinary differential equations with rational coefficients,}
 Physica D {\bf 2} (1981), 306-352.
\ss
[JM1]\ \  M. Jimbo and T. Miwa,
{\sl Monodromy preserving deformation of linear
ordinary differential equations with rational coefficients. II,}
 Physica D {\bf 2} (1981), 407-448.
\ss
[JM2]\ \  M. Jimbo and T. Miwa,
{\sl Monodromy preserving deformation of linear
ordinary differential equations with rational coefficients. III,}
 Physica D {\bf 4} (1983),  26-46.
\ss
[Ma]\ \ \ \  B. Malgrange, 
``\'Equations Differentielles \`a Coefficients Polynomiaux,''
 Birkha\"user,  Boston 1991.
\ss
[Mi]\ \ \ \  T. Miwa,
{\sl Painlev\'e property of monodromy preserving
equations and the analyticity of the $\tau$ function,}
 Publ. R.I.M.S. Kyoto University {\bf 17} (1981), 703-721.
\ss
[Mo1]\ \ \ \  G. Moore,
{\sl Geometry of the string equations,} 
 Comm. Math. Phys. {\bf 133} (1990), 261-304.
\ss
[Mo2]\ \ \ \  G. Moore,
{\sl Matrix models of 2D gravity and isomonodromic deformation,} 
Progress Theor. Phys., Suppl. no. 102 (1990), 255-285.
\ss
[Mt]\ \ \ \  T. F. Moutard, 
{\sl Note sur les \'equations differentielles lin\'eaires du second
ordre,}
C. R. Acad. Sci. Paris {\bf 80} (1876), 729.
\ss
[Sa]\ \ \ \  D. H. Sattinger, 
{\sl Hamiltonian hierarchies on semisimple Lie algebras,}
 Studies in Applied Math. {\bf 72} (1985), 65-86.
\ss
[Zu]\ \ \ \  V. Zurkowski, 
``Scattering for first order linear systems on the line
and B\"acklund transformations,''
Ph.D. Thesis, University of Minnesota - Minneapolis 1987.

\end